\documentclass[10pt,onecolumn,letterpaper]{article}
 \usepackage{cvpr}              % To produce the CAMERA-READY version
\usepackage[dvipsnames]{xcolor}

\definecolor{cvprblue}{rgb}{0.21,0.49,0.74}
\usepackage[pagebackref,breaklinks,colorlinks,citecolor=cvprblue]{hyperref}
\usepackage{multirow}

\def\paperID{*****} % *** Enter the Paper ID here
\def\confName{CVPR}
\def\confYear{2024}

\title{Proximal Gradient Descent Unfolding Dense-spatial Spectral-attention Transformer for Compressive Spectral Imaging}

\author{Ziyan Chen*\\
School of Automation, Guangdong University of Technology\\
{\tt\small zychen@gdut.edu.cn}
\and
Jing Cheng\\
School of Physics, South China University of Technology\\
{\tt\small phjcheng@scut.edu.cn}
}
\begin{document}
\maketitle
\abstract{ \rm{
 The Coded Aperture Snapshot Spectral Compressive Imaging (CASSI) system modulates three-dimensional hyperspectral images into two-dimensional compressed images in a single exposure. Subsequently, three-dimensional hyperspectral images (HSI) can be reconstructed from the two-dimensional compressed measurements using reconstruction algorithms. Among these methods, deep unfolding techniques have demonstrated excellent performance, with RDLUF-Mix$S^2$ achieving the best reconstruction results. However, RDLUF-Mix$S^2$ requires extensive training time, taking approximately 14 days to train RDLUF-Mix$S^2$-9stg on a single RTX 3090 GPU, making it computationally expensive. Furthermore, RDLUF-Mix$S^2$ performs poorly on real data, resulting in significant artifacts in the reconstructed images. In this study, we introduce the Dense-spatial Spectral-attention Transformer (DST) into the Proximal Gradient Descent Unfolding Framework (PGDUF), creating a novel approach called Proximal Gradient Descent Unfolding Dense-spatial Spectral-attention Transformer (PGDUDST). Compared to RDLUF-Mix$S^2$, PGDUDST not only surpasses the network reconstruction performance limit of RDLUF-Mix$S^2$ but also achieves faster convergence. PGDUDST requires only 58\% of the training time of RDLUF-Mix$S^2$-9stg to achieve comparable reconstruction results. Additionally, PGDUDST significantly alleviates the artifact issues caused by RDLUF-Mix$S^2$ in real experimental data, demonstrating superior performance and producing clearer reconstructed images.}
}

%\keyword{CASSI; Hyperspectral Image Reconstruction; Deep Learning; Signal Processing } 
\section{Introduction} 
\rm {Hyperspectral images (HSIs), consisting of numerous continuous and narrow spectral bands, provide a more comprehensive depiction of the observed scene compared to standard RGB images. Leveraging their inherently rich and detailed spectral characteristics, HSIs find extensive use in various computer vision tasks and graphical applications, including medical imaging \cite{HSImedicalimaging1, HSImedicalimaging2, HSImedicalimaging3}, image classification \cite{HSIimageclassification1, wambugu2021hyperspectral, imani2020overview}, remote sensing \cite{terentev2022current, peyghambari2021hyperspectral, gu2021multimodal, pande2023application}, object tracking \cite{li2020bae, zhao2023tmtnet, xiong2020material, tang2022robust}, and more. To obtain HSIs, conventional imaging systems equipped with spectrometers scan scenes in either the spatial or spectral dimension, often requiring a considerable amount of time \cite{gao2015optical}. Consequently, these traditional imaging systems fail to capture and measure dynamic scenes. Recently, snapshot compressive imaging (SCI) systems \cite{yuan2021snapshot, llull2013coded, wagadarikar2008single, wagadarikar2009video, chen2022snapshot, cao2016computational, du2009prism} have been developed to capture HSIs. These SCI systems modulate the three-dimensional (3D) hyperspectral image along the spectral dimension into a single two-dimensional (2D) compressive image. Among these SCI systems, coded aperture snapshot spectral imaging (CASSI) \cite{arce2013compressive, huang2021deep, gehm2007single, tsa, wang2015high} stands out due to its remarkable performance. CASSI acquires a spectral image by employing a coded aperture and dispersive components to modulate the spectral content within the scene. By capturing a 2D compressed measurement of the 3D data cube, the CASSI technique offers an efficient approach to acquiring spectral data. Nevertheless, reconstructing the 3D HSI cube from the 2D measurements presents a fundamental challenge for the CASSI system.

Based on CASSI, various reconstruction techniques can be categorized into four groups: model-based techniques \cite{wagadarikar2008single, kittle2010multiframe, liu2018rank, wang2016adaptive, zhang2019computational, tan2015compressive, figueiredo2007gradient}, Plug-and-play (PnP) algorithms \cite{chan2016plug, qiao2020snapshot, yuan2020plug, meng2021self, yuan2021plug}, end-to-end (E2E) approaches \cite{meng2020snapshot, tsa, lambda, hdnet, mst, cst}, and deep unfolding methods \cite{wang2020dnu, wang2019hyperspectral, dgsmp, fu2021bidirectional, zhang2022herosnet, admm, gap, dauhst, rdluf}. Model-based techniques rely on manually defined priors and assumptions, leading to limited generality and slow reconstruction speed. PnP algorithms integrate pre-trained denoising networks into traditional model-based methods, but the fixed nature of the pre-trained networks limits their performance. E2E methods learn the mapping function from measurements to HSIs using deep learning, but they often lack theoretical properties, interpretability, and flexibility due to differences in hardware systems. Deep unfolding methods use a multi-stage network to map measurements to the HSI cube, providing interpretability and deep learning capabilities, making them promising for HSI reconstruction. Among these methods, deep unfolding methods have shown superior performance by converting traditional iterative optimization algorithms into a series of deep neural network (DNN) blocks.

Among these deep unfolding methods, the reconstruction quality of RDLUF-Mix$S^2$\cite{rdluf} stands out as the best. RDLUF-Mix$S^2$\cite{rdluf} consists of two main modules: the Residual Degradation Learning Unfolding Framework (RDLUF), which integrates the residual degradation learning strategy into the data subproblem of Proximal Gradient Descent (PGD)\cite{beck2009fast}, and the Mixing priors across Spatial and Spectral (Mix$S^2$) Transformer, integrated into RDLUF as the denoiser for the prior subproblem. While RDLUF-Mix$S^2$ exhibits the best reconstruction performance among these approaches, boasting an impressive average PSNR of 39.57 dB for reconstructed images, it faces the challenge of longer training time. Achieving optimal reconstruction results with RDLUF-Mix$S^2$ requires approximately 14 days of training on a single RTX 3090 GPU, making it computationally expensive. Furthermore, RDLUF-Mix$S^2$ shows suboptimal performance with real experimental data, leading to reconstructed images containing significant artifacts.

In this paper, we introduce a novel Dense-spatial Spectral-attention Transformer (DST) and integrate it into the Proximal Gradient Descent Unfolding Framework (PGDUF), presenting a new method named Proximal Gradient Descent Unfolding Dense-spatial Spectral-attention Transformer (PGDUDST). PGDUF combines a deep unfolding framework with proximal gradient descent algorithms, gradually approximating the original high-dimensional data through multiple iterative stages. Each stage comprises a gradient descent module (primarily used to estimate residuals between the sensing matrix and the degradation matrix) and a proximal mapping module (DST), with the goal of enhancing adaptability to changes in the sensing matrix and improving reconstruction performance through multi-stage approximation. The proposed DST incorporates parallel-designed mechanisms, including Dense-spatial and Spectral-attention, enhancing the model's ability to capture complex details and structures. To further strengthen modeling capabilities in both spectral and spatial dimensions, a mechanism for bidirectional interaction of information in spatial and spectral domains is introduced. Additionally, the Dense-spatial mechanism combines Lightweight Inception and DenseBlock in a novel way, leveraging the strengths of both architectures. This combination excels in feature extraction, information aggregation, and gradient flow, thereby enhancing the overall performance of the model and achieving rapid convergence. Compared to RDLUF-Mix$S^2$, our proposed method, PGDUDST, exhibits three key advantages:

\begin{itemize}
\item[(1)] PGDUDST achieves reconstruction results comparable to RDLUF-Mix$S^2$-9stg but with a training time that is only 58\% of RDLUF-Mix$S^2$-9stg, resulting in nearly halved training time.
\item[(2)] PGDUDST surpasses the network reconstruction limit of RDLUF-Mix$S^2$.
\item[(3)] PGDUDST demonstrates superior performance with real experimental data, effectively mitigating the prominent artifacts found in RDLUF-Mix$S^2$ and yielding a cleaner and more detailed reconstruction in real-world experimental scenarios.
\end{itemize}

\section{The CASSI system}

\begin{figure*}[htpb]
\centering
\includegraphics[width=12cm,angle=0]{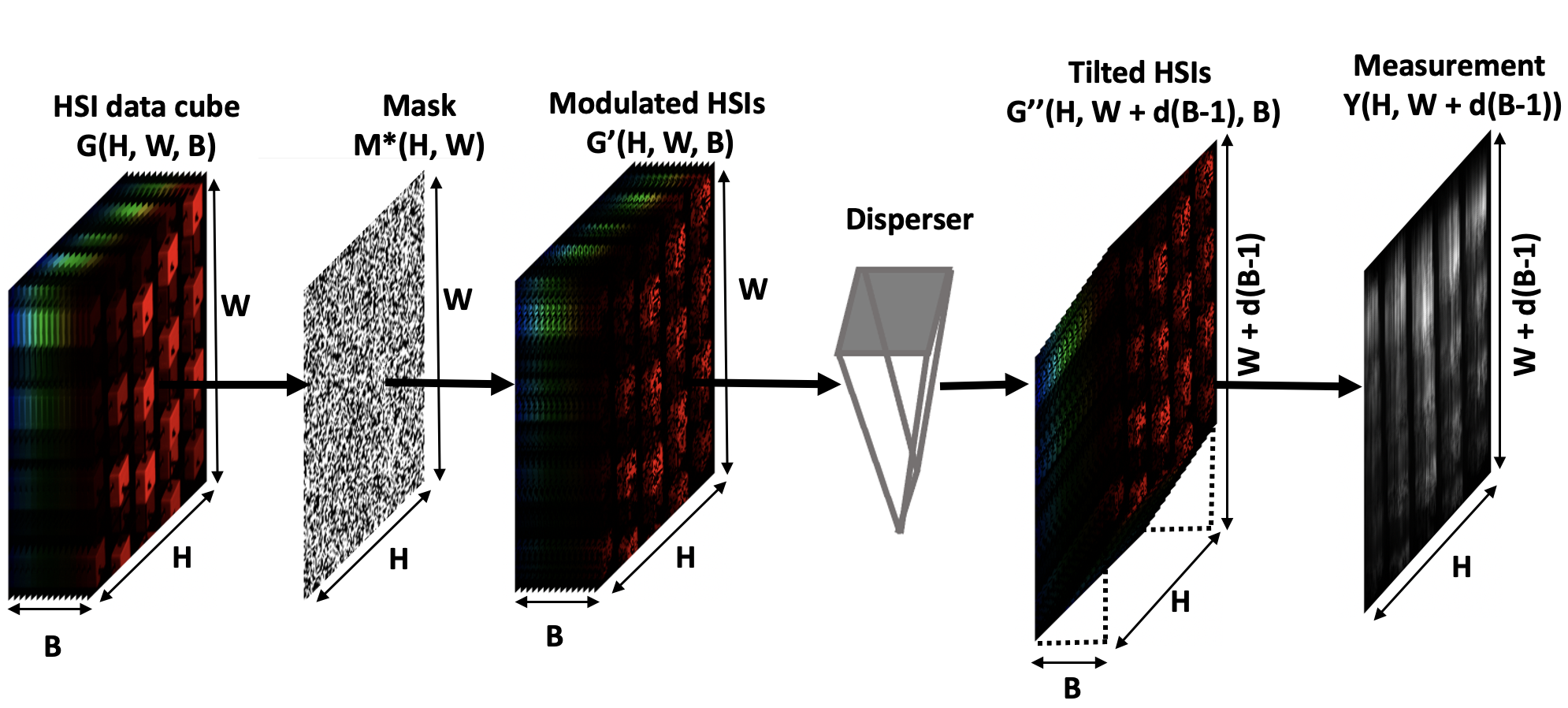}
\caption{The schematic of the CASSI system. }
\label{f1}
\end{figure*}

A concise CASSI system is illustrated in Fig.\ref{f1}. The 3D HSI data cube is denoted as ${\bf G} \in \mathbb{R}^{H \times W \times B}$, where $H$, $W$, and $B$ represent the height, width, and number of wavelengths of the 3D HSI, respectively. A physical mask ${\bf M^{}} \in \mathbb{R}^{H \times W}$ is utilized to modulate the 3D HSI ${\bf G}$. Each channel of this 3D HSI ${\bf G}$ is modulated by this physical mask, and this process can be expressed in mathematical formulas as follows:
\begin{eqnarray}
{\bf G'}(:, :, b) = {\bf G}(:, :, b) \odot {\bf M^{}}
\label{eq1}
\end{eqnarray}
where  $\odot$ represents element-wise multiplication, $b \in [1,2,3,..., B]$, and ${\bf G'} \in \mathbb{R}^{H \times W \times B}$ is the modulated 3D HSI. Consequently, the modulated 3D HSI ${\bf G'}$ is shifted by the disperser, expressed as:
\begin{eqnarray}
{\bf G''}(h, w, b) = {\bf G'}(h, w+d(b-1), b)
\label{eq2}
\end{eqnarray}
Here, ${\bf G''} \in \mathbb{R}^{H \times (W+d(B-1)) \times B}$ represents the tilted 3D HSI, which is shifted from ${\bf G'}$, and $d$ represents the shifting step. Finally, the 2D compressed measurement ${\bf Y} \in \mathbb{R} ^{H \times [W+d(B-1)]}$ can be obtained by
\begin{eqnarray}
{\bf Y}= \sum_{b=1}^{B} {\bf G''}(:, :, b) + {\bf N}
\label{eq3}
\end{eqnarray}
where ${\bf N} \in \mathbb{R} ^{H \times [W+d(B-1)]}$ represents the noise of the CASSI system.

Meanwhile, the degradation model of the CASSI system can be formulated as
\begin{eqnarray}
{y} = A {g} + n,
\label{eq4}
\end{eqnarray}
where ${g} \in \mathbb{R} ^{HWB \times 1}$ is reshaped from the 3D HSI ${\bf G} \in \mathbb{R} ^{H \times W \times B}$, ${y} \in \mathbb{R} ^{H[W+d(B-1)] \times 1}$ is reshaped from the modulated imaging speckle patterns ${\bf Y} \in \mathbb{R} ^{H \times [W+d(B-1)]}$ captured by the camera, $n \in \mathbb{R} ^{H[W+d(N-1)] \times 1}$ is reshaped from the noise $\bf N$ of the CASSI system. $A \in \mathbb{R} ^{H[W+d(B-1)] \times HWB}$ represents the sensing matrix calibrated based on pre-design.

\section{Method}
\subsection{The Proximal Gradient Descent algorithm}
Reconstructing the high-quality image \(g\) from the compressed measurement \(y\) is typically an ill-posed problem. Based on the maximizing the posterior probability (MAP) theory, Eq.\ref{eq4} can be solved as

\begin{eqnarray}
\hat{g} = \mathop{\rm argmin}\limits_g \frac{1}{2} \parallel y - Ag \parallel^{2} + \tau J(g),
\label{eq5}
\end{eqnarray}
where \(\frac{1}{2} \parallel y - Ag \parallel^{2}\) represents the data fidelity term, \(J(g)\) corresponds to the image prior term, and \(\tau\) is a hyperparameter that balances their relative importance. The PGD algorithm approximately formulates Eq.\ref{eq5} as an iterative convergence problem using the following iterative function:

\begin{eqnarray}
\textcolor{red}{\hat{g}^k = \mathop{\rm argmin}\limits_g \frac{1}{2\beta}  \parallel g - }\textcolor{blue}{\left(  \hat{g}^{k-1} - \beta A^{T}(A \hat{g}^{k-1} - y )       \right) }   \textcolor{red}{\parallel^{2} + \tau J(g)},
\label{eq6}
\end{eqnarray}
where \({\hat{g}^k}\) is the output of the \(k\)-th iteration, and \(\beta\) represents the step size.

Mathematically, the provided function consists of two main components: the blue part involves a gradient descent operation, and the red part can be addressed through the proximal operator \({\rm proj }_{\tau, J}\). This formulation results in a data subproblem and a prior subproblem, specifically gradient descent (Eq.\ref{eq7a}) and proximal mapping (Eq.\ref{eq7b}):

\begin{subequations}
\begin{eqnarray}
&&\textcolor{blue}{\upsilon^{k} = \hat{g}^{k-1} - \beta A^{T}(A \hat{g}^{k-1} - y )} \label{eq7a} \\
&&\textcolor{red}{{\hat{g}^k} = {\rm proj }_{\tau, J}(\upsilon^{k})} \label{eq7b}
\end{eqnarray}
\end{subequations}

The PGD algorithm iteratively updates \(\upsilon^{k}\) and \({\hat{g}^k}\) until convergence.

\subsection{Proximal Gradient Descent Unfolding Framework}

\begin{figure*}[htpb]
\centering
\includegraphics[width=15cm,angle=0]{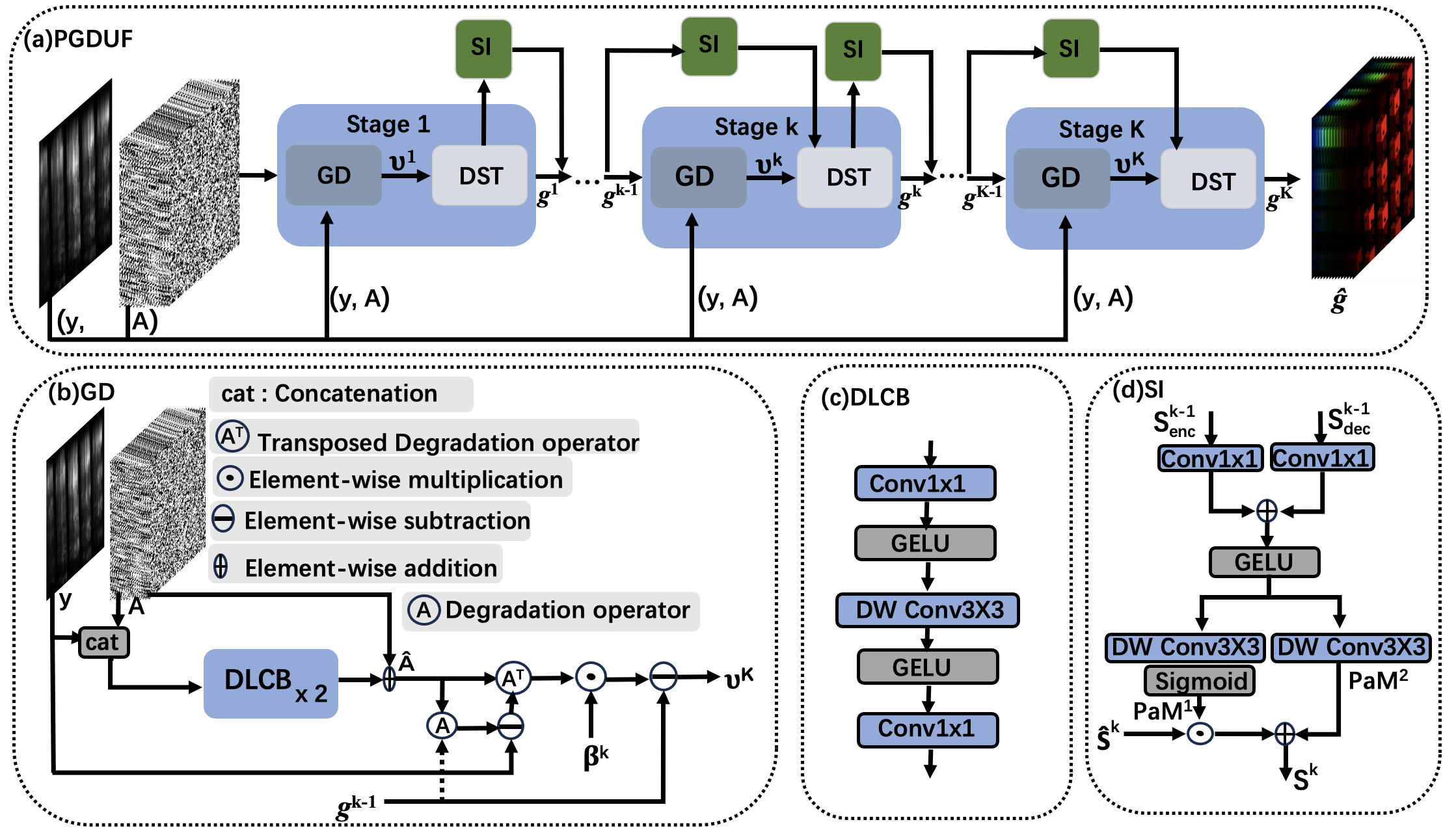}
\caption{(a)The architecture of the proposed PGDUF with K stages (iterations). (b) Gradient Descent module (GD). (c) Degradation Learning Convolution Block (DLCB). (d)Stage Interaction module (SI). }
\label{pgduf}
\end{figure*}
%(corresponding to the Dense-Spatial and Spectral-attention Transformer, DST)

Our proposed framework, PGDUF, is illustrated in Fig. \ref{pgduf}, representing a deep unfolding framework based on the Proximal Gradient Descent algorithm. PGDUF comprises several iteratively applied stages, each featuring a Gradient Descent module and a Proximal Mapping module. These modules correspond to the gradient descent (Eq.\ref{eq7a}) and proximal mapping (Eq.\ref{eq7b}) steps within a single iteration of the PGD algorithm. From a Bayesian perspective, the PM module can be interpreted as addressing a denoising problem \cite{chan2016plug, yuan2020plug}. As depicted in Fig. \ref{pgduf}(a), we have developed the Dense-spatial Spectral-attention Transformer (DST) as the PM module, and further details about DST will be illustrated in the upcoming subsection. Additionally, between each pair of adjacent stages, stage interaction has been introduced to enhance features and stabilize optimization through spatial adaptive normalization.

{\bf Gradient Descent module.} 
The Gradient Descent module employed in this work follows the one in the study \cite{rdluf}. This module estimates the residual between the sensing matrix $A$ and the degradation matrix $\hat{A}$, and its architecture is depicted in Fig. \ref{pgduf}(b). The GD module comprises two inputs: one is the compressed measurement $y$, and the other is the sensing matrix $A$. It consists of two Degradation Learning Convolution Blocks (DLCBs), whose architecture is also presented in Fig. \ref{pgduf}(c). The gradient descent step in the GD module can be expressed as follows:
\begin{eqnarray}
&& \upsilon^{k}=\hat{g}^{k-1} -\beta^{k} {\hat{A}}^{T^{k}}({\hat{A}^{k}} \hat{g}^{k-1} -y )  %\\ \nonumber
 \label{eq9}
\end{eqnarray}
in which $\beta$ is a learnable parameter, $k$ represents the number of stages, and the degradation matrix $\hat{A}$ can be calculated as $\hat{A}=A+{\rm T}_{\rm DLCB}( {\rm Concat}(y, A))$, 
where $\rm T_{DLCB}$ represents two cascaded DLCBs, and $\rm Concat$ denotes concatenation.

{\bf Stage Interaction module.} 
The stage interaction module employs a Spatial Adaptive Normalization (SPAN) approach to normalize features at the current stage\cite{SPAN1,SPAN2,rdluf}. SPAN utilizes features from the previous stage to generate modulation parameters, and the computation of modulation parameters is expressed as follows:

\begin{eqnarray}
&& \text{PaM}^{1}= \text{Sigmoid}\{\text{DWConv}[ \text{GELU} [\text{Conv}( \text{S}_{\text{enc}}^{k-1}) +\text{Conv}(\text{S}_{\text{dec}}^{k-1}) ] ]\} \\ \nonumber
&& \text{PaM}^{2}= \text{DWConv}[\text{GELU} [\text{Conv}( \text{S}_{\text{enc}}^{k-1}) +\text{Conv}(\text{S}_{\text{dec}}^{k-1}) ] ]%\\ \nonumber
\label{eq10}
\end{eqnarray}
Here, $k$ denotes the current stage number, $\text{PaM}^{1}$ ($\text{PaM}^{2}$) represents the generated modulation parameter of the current stage, and $\text{S}_{\text{enc}}^{k-1}$ ($\text{S}_{\text{dec}}^{k-1}$) denotes the encoder (decoder) features from the previous stage. Subsequently, the features of the current stage are modulated as follows:

\begin{eqnarray}
&&\text{S}^{k}= \text{PaM}^{1} \odot  \hat{\text{S}}^{k} + \text{PaM}^{2}  %\\ \nonumber
\label{eq10_1}
\end{eqnarray}     %, and $\odot$ signifies the operation of element-wise multiplication.
in which $\hat{\text{S}}^{k}$ denotes the intermediate output of the current stage. The proposed module for stage interaction boasts numerous advantages. To begin with, the network becomes more susceptible to vulnerabilities when information loss occurs due to the repetitive utilization of up and down sampling operations in the encoder-decoder. Secondly, the multi-scale features within each stage serve to enhance the features of the subsequent stage. Lastly, the optimization procedure for the network gains increased stability as it facilitates the smoother flow of information.

\subsection{Dense-spatial Spectral-attention Transformer}
  \begin{figure}[htpb]
\centering
\includegraphics[width=15cm,angle=0]{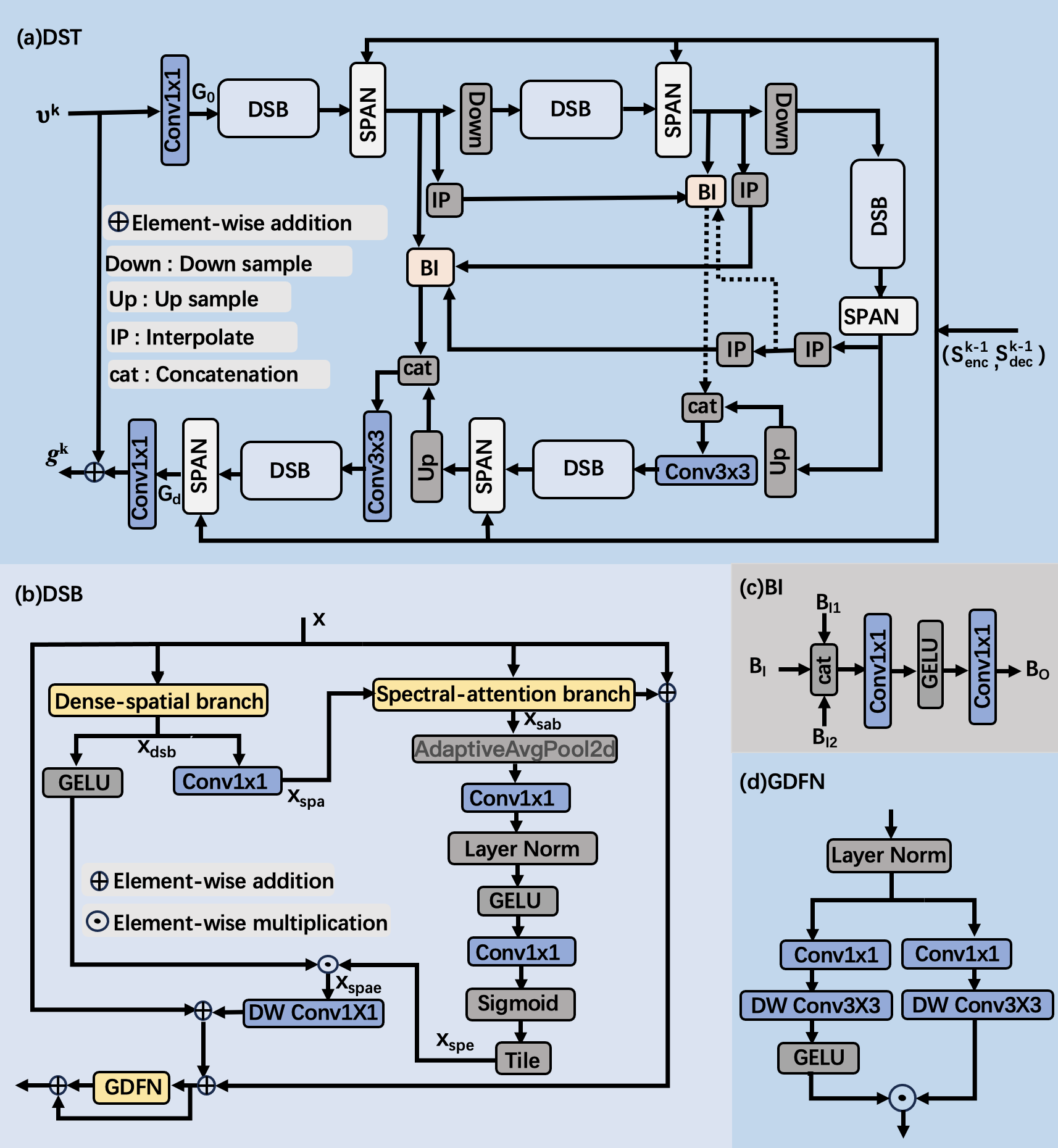}
\caption{(a)Dense-spatial Spectral-attention Transformer (DST). (b) Dense-spatial  Spectral-attention Block (DSB). (c) Block Interaction  (BI). (d) Gated-Dconv feed-forward network (GDFN).}
\label{dst}
\end{figure} 

   \begin{figure}[htpb]
\centering
\includegraphics[width=15cm,angle=0]{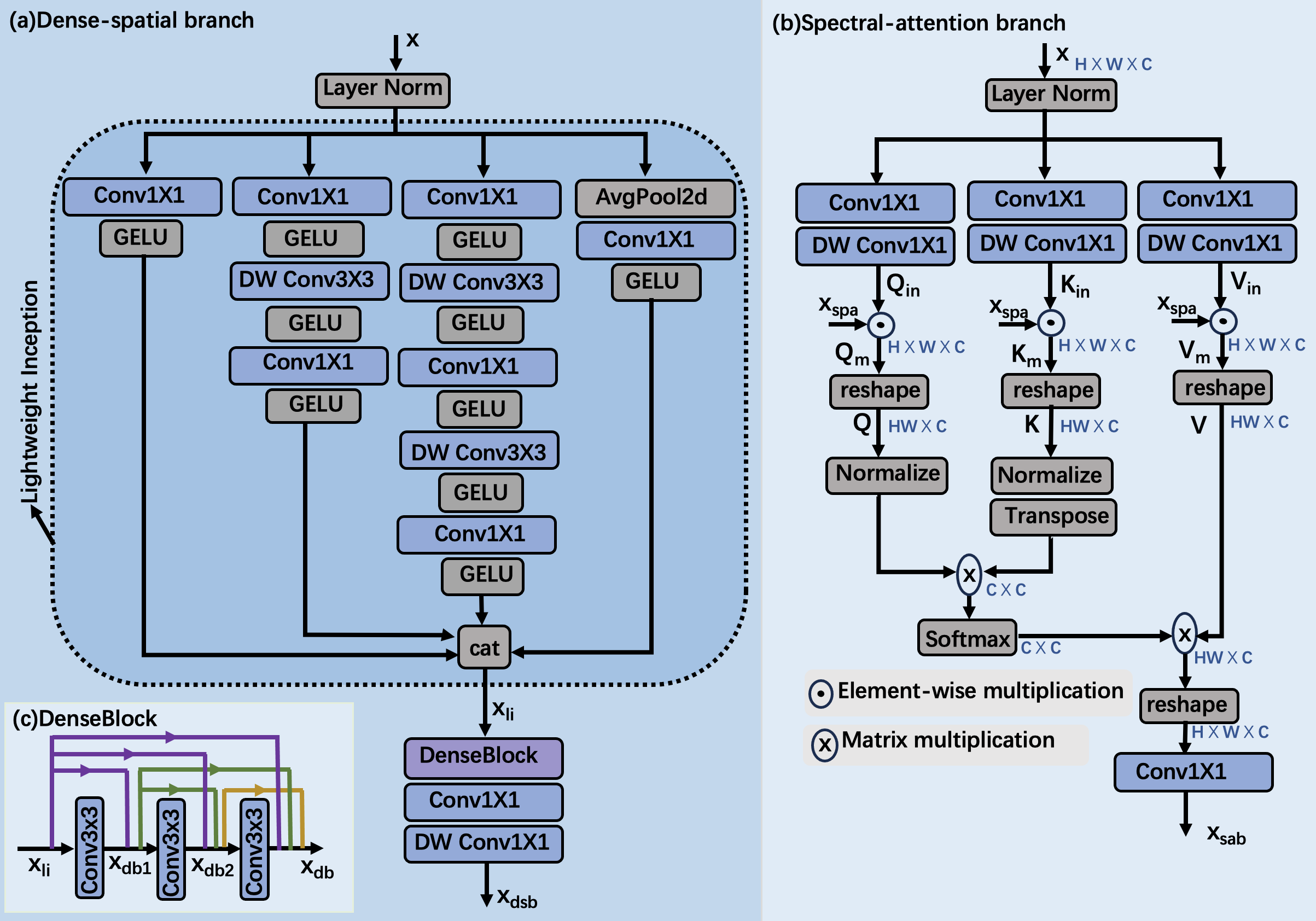}
\caption{(a)Dense-spatial Branch. (b)  Spectral-attention Branch. (c) DenseBlock.}
\label{dsb}
\end{figure}

In this section, we provide a comprehensive exposition of the proposed Dense-spatial Spectral-attention Transformer (DST).

  {\bf Network Architecture}. Illustrated in Fig.\ref{dst}(a), the DST employs a U-shaped structure with three hierarchical levels, comprising multiple foundational units referred to as Dense-spatial Spectral-attention blocks. To mitigate information loss resulting from up and down sampling operations, we introduce block interaction. This process entails the interpolation of features at different scales to a unified scale within each block interaction. Initially, the DST employs a Conv1$\times$1 operation to map $\upsilon^{k}$ to shallow features ${\rm G_0} \in \mathbb{R} ^{H \times \hat{W} \times B}$, where $\hat{W}=W+d(B-1)$. Subsequently, ${\rm G_0}$ traverses through all Dense-spatial Spectral-attention blocks and block interactions, becoming embedded in deep features ${\rm G_d} \in \mathbb{R} ^{H \times \hat{W} \times B}$. Finally, a Conv1 $\times$1 operation is applied to ${\rm G_d}$ to generate the denoised image $g_k$.

{\bf Block Interaction.} The Block Interaction module is depicted in Fig.\ref{dst}(c) and is designed to mitigate potential information loss that often occurs when resizing images or feature maps in the field of deep learning, particularly through common up-sampling and down-sampling operations. The Block Interaction module interpolates features at various scales, consolidating them within each block interaction. As a result, this process efficiently minimizes information loss during size transformations. The output of the Block Interaction module can be expressed as follows:
\begin{eqnarray}
&& \rm B_O= Conv\{GELU(Conv(Concat( B_{I}, B_{I1}, B_{I2})))\}%\\ \nonumber
\label{eq11}
\end{eqnarray}
where $\rm B_{Ii}(i=1, 2)$ represents features that have been interpolated to the same scale as $\rm B_{I}$.

  {\bf Dense-spatial Spectral-attention Block}.  The key component of the DST is the Dense-spatial Spectral-attention block, illustrated in Fig.\ref{dst}(b). This block comprises a Dense-spatial branch and a Spectral-attention branch, incorporating bidirectional interaction between spatial and spectral dimensions. Additionally, it includes a detailed gated-Dconv feed-forward network \cite{zamir2022restormer} as depicted in Fig.\ref{dst}(d). The details of the Dense-spatial branch, Spectral-attention branch, and the Spectral-Spatial Bidirectional Interaction are described as follows.  %
  
     {\bf Dense-spatial Branch}.  The Dense-spatial Branch is illustrated in Fig.\ref{dsb}(a). The employed Lightweight Inception module adheres to the design principles outlined in \cite{szegedy2017inception, szegedy2016rethinking}, incorporating a modification: the use of DWConv3 × 3 instead of Conv3 × 3, as depicted in Fig.\ref{dsb}(a) following \cite{rdluf}. Operating across various scales, the Lightweight Inception module processes visual information and aggregates it, allowing the subsequent layer to abstract features concurrently from different scales. This approach significantly enhances the network's capability to capture a broader spectrum of textures and details. Furthermore, the integration of the DenseBlock\cite{huang2017densely} module into the architecture, as depicted in Fig.\ref{dsb}(c), further enhances the network's efficacy. The DenseBlock process is articulated as follows:
\begin{eqnarray}
&&  \rm x_{db1}= \text{Concat}(x_{li}, \text{Conv}(x_{li})) \\ \nonumber
&&  \rm x_{db2}= \text{Concat}(x_{db1}, \text{Conv}(x_{db1}))\\  \nonumber
&&  \rm x_{db} = \text{Concat}(x_{db2},  \text{Conv}(\rm {x_{db2}}))
\label{eq12}
\end{eqnarray} 

Here, $\rm x_{li}$ represents the output of the Lightweight Inception module. DenseBlock structures facilitate feature reuse and promote gradient flow throughout the network, mitigating the vanishing gradient problem. This characteristic of DenseBlock modules is particularly advantageous in deep neural networks, enhancing the model's learning capacity and feature representation.

Following the Lightweight Inception module and the DenseBlock module, a sequence of operations, including Conv1 × 1 and DWConv1 × 1, is applied to generate the final output of the Dense-spatial Branch. The ultimate expression for the Dense-spatial Branch's output is defined as:
\begin{eqnarray}
&& { \rm x_{dsb}}= \text{DWConv}(\text{Conv}( \rm x_{db}))
\label{eq13}
\end{eqnarray} 
Here, $\rm x_{db}$ denotes the output of the DenseBlock module. The synergy of DenseBlock and Lightweight Inception harnesses the strengths of both architectures, resulting in a network that excels in feature extraction, information aggregation, and gradient flow, thereby enhancing the overall performance of the model and achieving rapid convergence.

 {\bf Spectral-attention branch.}The Spectral-attention branch follows the methodology outlined in \cite{zamir2022restormer}, with its key component being the Spectral-wise Multi-head Self-Attention module (S-MSA) \cite{mst}. Fig. \ref{dsb} (b) illustrates the Spectral-attention branch. The input $\rm x$ undergoes a Layer Norm layer to become $\rm x_{in}$, which is then embedded through Conv1 × 1 and DWConv3 × 3 layers. This embedding process results in $\rm Q_{in}=U_d^{Q}U_p^{Q}x_{in}$, $\rm K_{in}=U_d^{K}U_p^{K}x_{in}$, and $\rm V_{in}=U_d^{V}U_p^{V}x_{in}$, where $\rm U_p^{( )}$ denotes Conv1 × 1, and $\rm U_d^{( )}$ denotes DW Conv3 × 3. Subsequently, the obtained {\it query} ($\rm Q_{in}$), {\it key} ($\rm K_{in}$), and {\it value} ($\rm V_{in}$) are element-wise multiplied with $\rm x_{spa}$ to yield 
\begin{eqnarray}
%\nonumber
\label{eq14}
 \rm Q_m=Q_{in} \odot x_{spa} \\  \nonumber
 \rm K_m=K_{in} \odot x_{spa} \\   \nonumber
 \rm V_m=V_{in}\odot x_{spa}    \nonumber
\end{eqnarray} 

Then, $\rm Q_m, K_m$, and $\rm V_m$ are reshaped to obtain $\rm Q, K$, and $\rm V$, respectively, using the reshape operation, where $\rm Q \in \mathbb{R} ^{HW \times C}$, $\rm K \in \mathbb{R} ^{HW \times C}$, and $\rm V \in \mathbb{R} ^{HW \times C}$.

The final output of the Spectral-attention branch can be expressed as:

\begin{eqnarray}
&& { \rm x_{sab}}= \rm Conv(\text{Reshape}(\text{Attention}(\rm Q,K) \times \rm W)) 
\label{eq15}
\end{eqnarray} 

Here, $\rm Attention(Q,K)=\text{Softmax}\{\text{Normalize}(\rm Q) \times \text{Transpose}(\text{Normalize}(\rm K))\}$, and $\rm Attention(Q,K)$ $\in \mathbb{R} ^{C \times C}$.

   {\bf Spectral-Spatial Bidirectional Interaction.} Having familiarity with \cite{chen2022mixformer}, we introduce the Spectral-Spatial Bidirectional Interaction module. For spatial interaction, the output features $\rm x_{dsb}$ from the Dense-spatial Branch undergo Conv1 × 1 to generate a spatial attention map, denoted as $\rm x_{spa}$:

\begin{eqnarray}
\rm x_{spa} = \text{Conv}(\rm x_{dsb})
\label{eq16}
\end{eqnarray}    

This spatial attention map $\rm x_{spa}$ is then applied to the {\it query} ($\rm Q_{in}$),  {\it key} ($\rm K_{in}$), and  {\it value} ($\rm V_{in}$)   in a spatial attention manner, as described in Eq.\ref{eq14}.
   
For spectral interaction, the output features $\rm x_{sab}$ from the Spectral-attention branch undergo a series of operations \cite{hu2018squeeze} to obtain a spectral attention weight map, denoted as $\rm x_{spe}$:

\begin{eqnarray}
\rm x_{spe} = \text{Tile}(\text{Sigmoid}(\text{Conv}(\text{GELU}(\text{LN}(\text{Conv}(\text{AdaptiveAvgPool2d}(\rm x_{sab})))))))
\label{eq17}
\end{eqnarray} 

Here, $\rm LN$ represents the Layer Norm layer. Subsequently, this spectral attention weight map $\rm x_{spe}$ is applied to the output features $\rm x_{dsb}$ of the Dense-spatial Branch in a spectral attention manner, resulting in the generation of the Spectral-Spatial attention map $\rm x_{spae}$:

\begin{eqnarray}
\rm x_{spae} = \rm x_{spe} \odot \text{GELU}(\rm x_{dsb})
\label{eq18}
\end{eqnarray}

\section{ Experiments}

\begin{table*}[h!]
\caption{Comparisons of PSNR (upper entry in each cell) and SSIM (lower entry in each cell) for different methods across 10 simulation scenes (S1-S10).}
\label{table_1}
\renewcommand{\arraystretch}{1.5}.  %行宽
\setlength{\tabcolsep}{2.8pt}   %列宽
\centering
%\scriptsize
\footnotesize
%\tiny
\begin{tabular}{|c |c| c|c|c|c| c|c|c|c |c| c| c|c| c|c |c|c| c|c |c|c| c|c |c|c| c|} % {p{8cm} p{2.5cm}<{\centering}}
	\hline
	% \multirow{2}{*}{Name}，2为所占的行数，此语句可以使得内容垂直居中
	% \multicolumn{2}{c|}{Flag}，2为所占的列数，格式由第二个{}控制
	% \cline{2-3}指本行的2,3列画横线
	\multirow{1}{*}{\bf Algorithms}   
	%&\multicolumn{1}{c|}{\bf Params} 
	%&\multicolumn{1}{c|}{\bf GFLOPS}  
	&\multicolumn{1}{c|}{\bf S1}	 
	&\multicolumn{1}{c|}{\bf S2} 
	
	&\multicolumn{1}{c|}{\bf S3} 
	&\multicolumn{1}{c|}{\bf S4}  
	&\multicolumn{1}{c|}{\bf S5}	 
	%&\multicolumn{2}{c|}{GISCnet:32}
	
       &\multicolumn{1}{c|}{\bf S6} 
	&\multicolumn{1}{c|}{\bf S7}  
	&\multicolumn{1}{c|}{\bf S8}	 
	&\multicolumn{1}{c|}{\bf S9}	
	&\multicolumn{1}{c|}{\bf S10}	
	&\multicolumn{1}{c|}{\bf Avg}	
	
			\\ \cline{1-12}

	               %  \\  \hline  
\multirow{2}{*}{TwIST \cite{bioucas2007new} }    %&   \multirow{2}{*}{-}  &  \multirow{2}{*}{-}  
 & 25.16  & 23.02 & 21.40  & 30.19  & 21.41  &20.95  & 22.20  & 21.82  & 22.42  & 22.67 & 23.12 \\
& 0.700  & 0.604  &0.711  & 0.851  & 0.635  & 0.644  & 0.643 & 0.650  & 0.690  & 0.569  &0.669	       
	       \\  \cline{1-12}   
	       
\multirow{2}{*}{GAP-TV \cite{yuan2016generalized} }    % &   \multirow{2}{*}{-}  &  \multirow{2}{*}{-} 
 & 26.82  & 22.89 & 26.31  & 30.65  & 23.64  &21.85  & 23.76  & 21.98  & 22.63 & 23.10 & 24.36 \\
   & 0.754  & 0.610  &0.802 & 0.852  & 0.703  & 0.663  & 0.688 & 0.655  & 0.682  & 0.584  & 0.669	       
	       \\  \cline{1-12}  
	       
\multirow{2}{*}{DeSCI \cite{liu2018rank} }     %&   \multirow{2}{*}{-}  &  \multirow{2}{*}{-} 
  & 27.13  & 23.04 & 26.62  & 34.96  & 23.94  &22.38  & 24.45  & 22.03  & 24.56 & 23.59 & 25.27 \\
   & 0.748  & 0.620  &0.818 & 0.897  & 0.706  & 0.683  & 0.743 & 0.673  & 0.732  & 0.587 &0.721	       
	       \\  \cline{1-12}  	

            \multirow{2}{*}{$\lambda$-net \cite{lambda}}  % &   \multirow{2}{*}{62.64M}  &  \multirow{2}{*}{117.98}  
  & 30.10  & 28.49 & 27.73  & 37.01  & 26.19  & 28.64  & 26.47  & 26.09  & 27.50  & 27.13 & 28.53 \\
   & 0.849  & 0.805 & 0.870  & 0.934  & 0.817  & 0.853  & 0.806 & 0.831  & 0.826  & 0.816  &0.841	
					              \\  \cline{1-12} 
					                
   \multirow{2}{*}{TSA-Net \cite{tsa}}     %&   \multirow{2}{*}{44.25M}  &  \multirow{2}{*}{110.06} 
& 32.03 & 31.00 & 32.25  & 39.19  & 29.39  & 31.44  & 30.32  & 29.35  & 30.01  & 29.59 & 31.46 \\
     & 0.892  & 0.858  & 0.915  & 0.953  & 0.884  & 0.908  & 0.878 & 0.888  & 0.890  & 0.874  & 0.894		    
					              \\  \cline{1-12} 					              
					              
    \multirow{2}{*}{V-DUnet \cite{czydgi}}     %&   \multirow{2}{*}{21.24M}  &  \multirow{2}{*}{52.65} 
  & 32.70  & 32.21 & 32.91 & 39.01  & 30.10  & 31.45  & 31.00  & 29.78  & 31.62  & 29.64 & 32.04 \\
 & 0.892  & 0.891 & 0.933  & 0.956  & 0.910  & 0.914 & 0.885 & 0.900  & 0.897  & 0.873  &0.905 			     
					              \\  \cline{1-12} 	
					              
 \multirow{2}{*}{DGSMP \cite{dgsmp}}  %&   \multirow{2}{*}{3.76M}  &  \multirow{2}{*}{646.65}    
 & 33.26  & 32.09 & 33.06  & 40.54  & 28.86  & 33.08  & 30.74  & 31.35  & 31.66 & 31.44 & 32.63 \\
 & 0.915  & 0.898  & 0.925  & 0.964  & 0.882  & 0.937  & 0.886 & 0.923  & 0.911  & 0.925  & 0.917
 			
					              \\  \cline{1-12}		
					              
\multirow{2}{*}{GAP-Net \cite{gap}}  %&   \multirow{2}{*}{4.27M}  &  \multirow{2}{*}{78.58}    
 & 33.74 & 33.26 & 34.28  & 41.03  & 31.44  & 32.40  & 32.27  &30.46  & 33.51 & 30.24 & 33.26 \\
  & 0.911  & 0.900  & 0.929  & 0.967  & 0.919  & 0.925 & 0.902 & 0.905  & 0.915  & 0.895  & 0.917
 			
					              \\  \cline{1-12}	
 \multirow{2}{*}{ADMM-Net \cite{admm}}  %&   \multirow{2}{*}{4.27M}  &  \multirow{2}{*}{78.58}    
 & 34.12  & 33.62 & 35.04  & 41.15  & 31.82  & 32.54  & 32.42  & 30.74  & 33.75 & 30.68 & 33.58 \\
 & 0.918  & 0.906  & 0.931  & 0.966  & 0.922  & 0.924 & 0.896 & 0.907 & 0.915  & 0.895  & 0.918
 			
					              \\  \cline{1-12}

\multirow{2}{*}{HDnet \cite{hdnet}}     %&   \multirow{2}{*}{2.37M}  &  \multirow{2}{*}{154.76} 
   & 35.14  & 35.67 & 36.03  & 42.30  & 32.69  & 34.46  & 33.67  & 32.48  & 34.89  & 32.38 & 34.97 \\
   & 0.935  & 0.940  & 0.943  & 0.969  & 0.946  & 0.952  & 0.926 & 0.941 & 0.942  & 0.937  &0.943
 					             \\  \cline{1-12}

		 \multirow{2}{*}{MST-L \cite{mst}}   % &   \multirow{2}{*}{2.03M}  &  \multirow{2}{*}{28.15}  
 & 35.40  & 35.87 & 36.51  & 42.27  & 32.77  & 34.80 & 33.66  & 32.67  & 35.39  & 32.50 & 35.18 \\
  & 0.941  & 0.944 & 0.953  & 0.973  & 0.947  & 0.955  & 0.925 & 0.948  & 0.949  & 0.941 &0.948
					              \\  \cline{1-12} 	
					              
   \multirow{2}{*}{MST++ \cite{mst_plus_plus}}   %  &   \multirow{2}{*}{1.33M}  &  \multirow{2}{*}{19.42} 
 & 35.80  & 36.23 & 37.34  & 42.63  & 33.38  & 35.38 & 34.35  & 33.71  & 36.67 & 33.38 &35.99 \\
     & 0.943  & 0.947  & 0.957  & 0.973  & 0.952  & 0.957  & 0.934 & 0.953  & 0.953  & 0.945  &0.951

					              \\  \cline{1-12} 						              
					              
     \multirow{2}{*}{CST-L* \cite{cst}}    % &   \multirow{2}{*}{3.00M}  &  \multirow{2}{*}{40.01} 
      & 35.96  & 36.84 & 38.16  & 42.44  & 33.25  & 35.72  & 34.86  & 34.34  & 36.51  & 33.09 & 36.12 \\
  & 0.949  & 0.955  & 0.962  & 0.975  & 0.955  & 0.963  & 0.944 & 0.961  & 0.957  & 0.945  &0.957 
				     
					              \\  \cline{1-12} 					              
					               
%    \multirow{2}{*}{RDLUF-Mix$S^2$-5stg \cite{dauhst}}    &   \multirow{2}{*}{1.89M}  &  \multirow{2}{*}{127.43}  
%    \multirow{2}{*}{RDLUF-Mix$S^2$-7stg \cite{dauhst}}    &   \multirow{2}{*}{1.89M}  &  \multirow{2}{*}{179.26}  
					              
    \multirow{2}{*}{DAUHST-9stg \cite{dauhst}}   % &   \multirow{2}{*}{6.15M}  &  \multirow{2}{*}{79.50}  
 %      \multirow{2}{*}{RDLUF-Mix$S^2$-11stg \cite{dauhst}}    &   \multirow{2}{*}{1.89M}  &  \multirow{2}{*}{282.93}  
  & 37.25 & 39.02 & 41.05  & 46.15 & 35.80  & 37.08  & 37.57  & 35.10 & 40.02    & 34.59 & 38.36 \\
  & 0.958  & 0.967  & 0.971  & 0.983 & 0.969  & 0.970  & 0.963  & 0.966 & 0.970   & 0.956  &0.967 			
					              \\  \cline{1-12} 	   

   % \multirow{2}{*}{DAUHST-9stg-czy \cite{dauhst}}     
 %& 37.74 & 40.31 & 42.35  & 46.82 & 36.54  & 37.34  & 38.14  & 36.03 & 41.08    & 35.02 & \bf39.14 \\
 %& 0.962  & 0.974  & 0.976  & 0.986 & 0.973  & 0.973  & 0.967  & 0.971 & 0.975   & 0.960  &\bf 0.972 		
%					              \\  \cline{1-14} 	

 \multirow{2}{*}{RDLUF-MixS$^2$-5stg \cite{rdluf}}    %&   \multirow{2}{*}{1.89M}  &  \multirow{2}{*}{127.43} 
 	     & 37.30  & 39.39& 42.06  & 46.89  & 35.74 & 37.03  & 37.05  & 35.18  & 40.64  & 34.58 & 38.59 \\
    & 0.960  & 0.971  & 0.975  & 0.988  & 0.969  & 0.971  & 0.959 & 0.968  & 0.973  & 0.957  & 0.969	               
					             \\  \cline{1-12} 	
					             
  \multirow{2}{*}{RDLUF-MixS$^2$-7stg \cite{rdluf}}    %&   \multirow{2}{*}{1.89M}  &  \multirow{2}{*}{179.26} 
 & 37.65  & 40.45& 43.00  & 47.40  & 36.78 & 37.56  & 38.25  & 35.86  & 41.71  & 34.83 & 39.35 \\
   & 0.963  & 0.976  & 0.978  & 0.990  & 0.974  & 0.974  & 0.967 & 0.971  & 0.978  & 0.959  & 0.973	               
					             \\  \cline{1-12}

 \multirow{2}{*}{RDLUF-MixS$^2$-9stg \cite{rdluf}}    % &   \multirow{2}{*}{1.89M}  &  \multirow{2}{*}{231.09} 
 & \bf37.94  & 40.95 & 43.25  & 47.83  & 37.11  & 37.47  & 38.58  & 35.50  & 41.83  & \bf35.23 & 39.57 \\
   & 0.966  & 0.977  & 0.979  & 0.990  & 0.976  & 0.975  & 0.969 & 0.970  & 0.978  & \bf0.962  & 0.974	               			             
					             \\  \cline{1-12} 	

 %		 \multirow{2}{*}{Ours 3stage}     
 %& 37.02  & 38.81 & 41.29  & 45.97  & 34.80  & 36.48  & 36.45  & 34.62  & 39.86  & 33.72 & 37.90\\
 % & 0.957  & 0.967  & 0.973  & 0.986  & 0.965  & 0.969  & 0.955 & 0.965 & 0.969  & 0.951  & 0.966	               			             \\  \cline{1-14}	

   \multirow{2}{*}{Ours 5stage}       %&   \multirow{2}{*}{2.46M}  &  \multirow{2}{*}{153.60} 
& 37.49  & 40.22 & 42.68  & 47.30  & 36.11  & 37.12  & 38.08  & 35.46  & 41.38  & 34.99 & 39.08\\
    & 0.964  & 0.975  & 0.978  & 0.988  & 0.972  & 0.972  & 0.966 & 0.969 & 0.977  & 0.961  & 0.972	               			             \\  \cline{1-12}

 \multirow{2}{*}{Ours 7stage}      % &   \multirow{2}{*}{2.46M}  &  \multirow{2}{*}{215.90} 
  & 37.89  & 40.90 & 43.34  & 47.64  & 36.60  & 37.46  & 38.46  & 36.14  & 42.28  & 35.15 & 39.59\\
 & 0.965  & 0.978  & 0.979  & 0.990  & 0.976  & 0.975  & 0.969 & 0.974  & 0.979  & 0.962  & 0.974	               			             \\  \cline{1-12}
   						             
%\multirow{2}{*}{Ours 11stage }       &   \multirow{2}{*}{2.46M}  &  \multirow{2}{*}{340.5} 					             					             
 \multirow{2}{*}{Ours 9stage }      % &   \multirow{2}{*}{2.46M}  &  \multirow{2}{*}{278.20} 
 & 37.90  & \bf41.22 & \bf43.55  & \bf47.87  & \bf37.25  & \bf37.62  & \bf38.92  & \bf36.28 & \bf42.53 & 35.06 & \bf39.82 \\
  & \bf0.966  & \bf0.979  &\bf 0.980  & \bf0.990  & \bf0.977  & \bf0.975  & \bf0.969 & \bf0.973  & \bf0.980  & 0.961  & \bf0.975	               
  
  %          \\  \cline{1-12} 	
          
%                \multirow{2}{*}{DMST++}    		                                      				 				
 %& 26.64  & 30.03 & 30.40  & 27.89  & 33.79  & 25.92  & 29.48  & 32.69  & 26.69  & 26.43 & \bf29.00 \\
%  & 0.774  & 0.865  & 0.829  & 0.861  & 0.885  & 0.891  & 0.908 & 0.914  & 0.806  & 0.859  & \bf0.859 			             
					               %\cline{1-10}                 
             \\  \hline    		                                   
\end{tabular}
\end{table*}

\subsection{Experimental Settings}

Following the configurations of TSA-Net \cite{tsa}, we utilize 28 wavelengths ranging from 450 nm to 650 nm, obtained through spectral interpolation manipulation, for HSIs. Our experiments cover both simulated and real HSI datasets.

%PSNR\cite{PSNR} and structural similarity (SSIM)\cite{SSIM} are used to evaluate the performance of all methods.
{\bf Simulation HSI Data}.We utilize two simulated HSI datasets, CAVE \cite{cave} and KAIST \cite{kaist}. The CAVE dataset comprises 32 HSIs with a spatial size of 512×512, while the KAIST dataset includes 30 HSIs with a spatial size of 2704×3376. Following the methodology of TSA-Net \cite{tsa}, we designate CAVE as the training set. For testing purposes, we select 10 scenes from the KAIST dataset.

{\bf Real HSI Data}. We employ the real HSI dataset obtained through the CASSI system, as developed in TSA-Net \cite{tsa}.

{\bf Evaluation Metrics}. We choose peak signal-to-noise ratio (PSNR) \cite{PSNR} and structural similarity (SSIM) \cite{SSIM} as the evaluation metrics to assess the performance of HSI reconstruction.

{\bf Implementation Details}. The PGDUDST model was implemented using the PyTorch framework and trained with the Adam optimizer, utilizing hyperparameters $\beta_1 = 0.9$ and $\beta_2 = 0.999$. The training process extended over 200 epochs, employing the cosine annealing scheduler with linear warm-up. The learning rate and batch size were set to $2 \times 10^{-4}$ and 1, respectively. When performing experiments on simulated data, the networks were fed patches with a spatial size of 256×256 cropped from the 3D cubes. In the case of real HSI reconstruction, the patch size was adjusted to 660×660 to align with real-world measurements. The dispersion shift steps were configured as 2, and data augmentation techniques, including random flipping and rotation, were applied. The model aimed to minimize the Charbonnier loss.

\subsection{Quantitative Results}
In our study, we conducted a comprehensive comparative analysis of the proposed PGDUDST method and state-of-the-art (SOTA) HSI restoration techniques. We compared the results of PGDUDST with 15 SOTA methods, including three model-based methods (TwIST \cite{bioucas2007new}, GAP-TV \cite{yuan2016generalized}, and DeSCI \cite{liu2018rank}), seven end-to-end methods (V-DUnet \cite{czydgi}, $\lambda$-net \cite{lambda}, TSA-net \cite{tsa}, HDNet \cite{hdnet}, MST \cite{mst}, CST \cite{cst}, and MST++ \cite{mst_plus_plus}), and five deep unfolding methods (ADMM-net \cite{admm}, GAP-net \cite{gap}, DGSMP \cite{dgsmp}, DAUHST \cite{dauhst}, and RDLUF-Mix$S^2$ \cite{rdluf}) across 10 simulation scenes. All techniques were trained using the same datasets and evaluated under identical settings as DGSMP \cite{dgsmp} to ensure fair comparisons.

The effectiveness of different methods was assessed based on the metrics of peak signal-to-noise ratio (PSNR) and structural similarity index (SSIM). The corresponding results for the 10 simulated scenes are presented in Table \ref{table_1}. From Table \ref{table_1}, it is evident that PGDUDST exhibits the best reconstruction performance, surpassing other approaches. Compared to $\lambda$-net \cite{lambda}, TSA-net \cite{tsa}, V-DUnet \cite{czydgi}, GAP-net \cite{gap}, HDNet \cite{hdnet}, CST-L-plus \cite{cst}, and RDLUF-Mix$S^2$-9stg \cite{rdluf}, the proposed method with 9 stages achieves significant improvements, with average gains of 11.29 dB, 8.36 dB, 7.78 dB, 6.56 dB, 4.85 dB, 3.7 dB, and 0.23 dB, respectively.

What stands out is that, although our network's average PSNR for reconstruction seems to be only 0.23 dB higher than that of RDLUF-Mix$S^2$-9stg \cite{rdluf}, we wish to emphasize that our network achieves comparable reconstruction results to RDLUF-Mix$S^2$-9stg \cite{rdluf} while requiring only 58\% of the training time. Additionally, it is crucial to highlight that our approach surpasses the network reconstruction performance limit of RDLUF-Mix$S^2$. Detailed explanations for these two aspects will be provided in the following sections, namely "Comparison of Training Times with RDLUF-Mix$S^2$" and "Comparative Analysis of Network Reconstruction Performance Limits with RDLUF-Mix$S^2$."

 { \bf Comparison of Training Times with RDLUF-Mix$S^2$. }  Considering that the original configuration of RDLUF-Mix$S^2$ \cite{rdluf} specified 300 epochs, we opted to train our model for an equivalent number of epochs. This decision facilitates a more intuitive comparison of the reconstruction PSNR and SSIM between the two methods. Furthermore, it ensures a fair evaluation of training duration, enhancing the comparability of our visualization results. It is important to emphasize, however, that our approach only requires setting epochs to 200.

     \begin{figure}[htpb]
\centering
\includegraphics[width=14cm,angle=0]{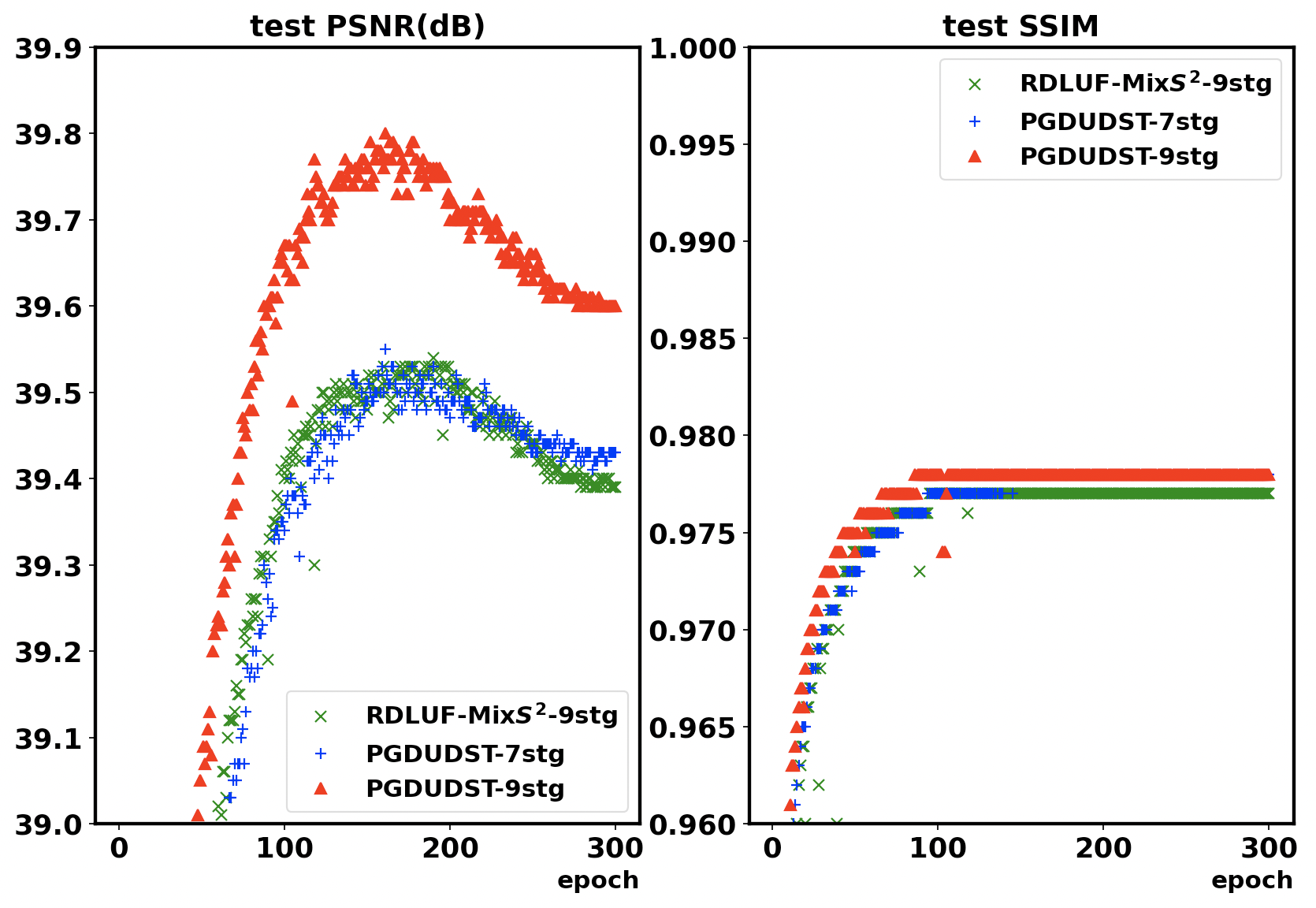}
\caption{Comparison of PSNR and SSIM for the models RDLUF-MixS$^2$-9stage, PGDUDST-7stage, and PGDUDST-9stage on 10 simulated test images.}
\label{psnr_ssim3}
\end{figure}

The reason our PGDUDST only requires 200 epochs, instead of being set to 300 epochs like RDLUF-Mix$S^2$ \cite{rdluf}, can be found in Figure \ref{psnr_ssim3}. As depicted in the figure, our PGDUDST, including both PGDUDST-7stg and PGDUDST-9stg, converges to the best PSNR value around epoch 150. In contrast, RDLUF-Mix$S^2$ \cite{rdluf} reaches its best PSNR value around epoch 200. This indicates that the convergence speed of our PGDUDST is much faster than that of RDLUF-Mix$S^2$ \cite{rdluf}. Consequently, we can significantly reduce the number of training epochs to 200. Additionally, since RDLUF-Mix$S^2$ \cite{rdluf} converges to its best PSNR value around epoch 200, setting the epochs for RDLUF-Mix$S^2$ \cite{rdluf} to 200 might be risky, as it may not converge to the best reconstruction result within 200 epochs.

From Figure \ref{psnr_ssim3}, we observe that our PGDUDST-7stg can achieve reconstruction results comparable to those of RDLUF-Mix$S^2$-9stg. Strictly speaking, it even outperforms RDLUF-Mix$S^2$-9stg. Consequently, we conducted a detailed comparison of the training duration required for PGDUDST-7stg (which can achieve reconstruction results comparable to RDLUF-Mix$S^2$-9stg) and RDLUF-Mix$S^2$-9stg (which achieved the best reconstruction results of RDLUF-Mix$S^2$). RDLUF-Mix$S^2$-9stg requires a total training time of 339.3 hours for 300 epochs, approximately 14.1 days. In contrast, PGDUDST-7stg requires 293.5 hours for 300 epochs, approximately 12.2 days. Even with the training epochs set to 300, PGDUDST-7stg still outperforms RDLUF-Mix$S^2$-9stg in terms of time efficiency, saving approximately 14\% of the total training time while maintaining comparable results to RDLUF-Mix$S^2$-9stg.

Due to the rapid convergence of our network, setting the epochs to 300 is unnecessary; we believe that setting PGDUDST's epochs to 200 is sufficient. PGDUDST-7stg requires 197.7 hours for 200 epochs, approximately 8.2 days. This results in a 42\% reduction in our network's training time. In summary, PGDUDST-7stg requires only 58\% of the training time compared to RDLUF-Mix$S^2$-9stg, yet achieves comparable training reconstruction results. Moreover, as evident from Figure \ref{psnr_ssim3}, PGDUDST-9stg outperforms RDLUF-Mix$S^2$-9stg in training results. The training time required for 200 epochs of PGDUDST-9stg is only 256.1 hours, approximately 10.7 days. Compared to RDLUF-Mix$S^2$-9stg, this represents a 25\% reduction in training time, indicating that PGDUDST can achieve superior results in a shorter training period than RDLUF-Mix$S^2$-9stg.

In summary, PGDUDST significantly shortens training time through two key mechanisms. Firstly, by alleviating the computational burden on the network, we achieve a notable reduction in training time (even with 300 epochs of training, PGDUDST-7stg requires 14\% less time than RDLUF-Mix$S^2$-9stg). Secondly, by accelerating the convergence speed of the network, we further minimize training duration. PGDUDST's convergence speed surpasses that of RDLUF-Mix$S^2$ \cite{rdluf}, enabling us to reduce the number of training epochs and thus significantly reduce overall training time. In total, PGDUDST only requires 58\% of the training time needed by RDLUF-Mix$S^2$-9stg to achieve comparable reconstruction results.

 \begin{figure}[htpb]
\centering
\includegraphics[width=17.5cm,angle=0]{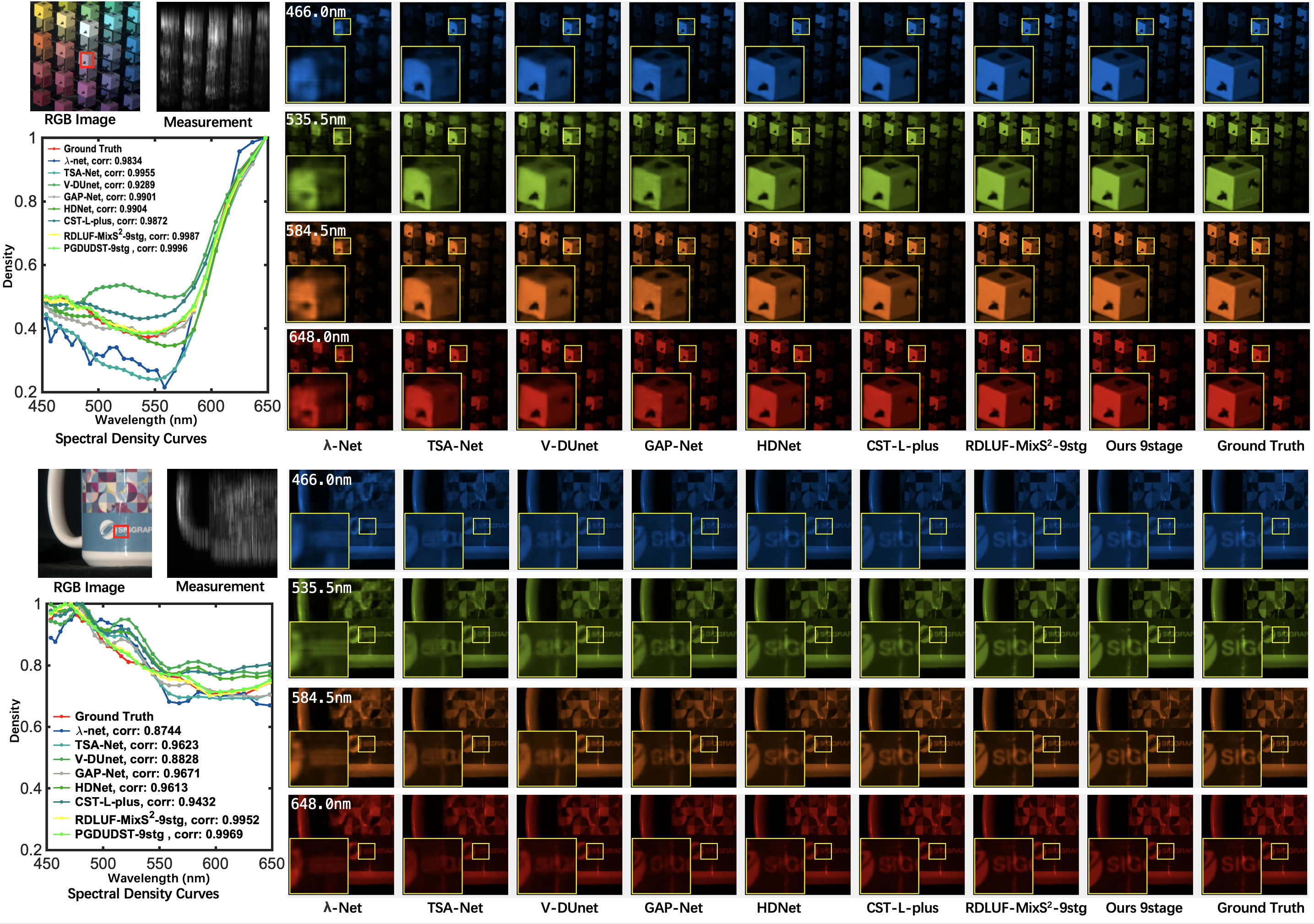}
\caption{Comparisons of reconstructed simulation HSI for Scene 2 (top) and Scene 5 (bottom) using 4 out of 28 spectral channels are presented. The evaluation includes 7 SOTA algorithms alongside our PGDUDST-9stg method. The spectral curves correspond to the selected red box in the RGB image. Zoom in for a more detailed examination.}
\label{simu_spectral}
\end{figure}

\begin{table*}
\caption{Comparison of PSNR (upper entry in each cell) and SSIM (lower entry in each cell) across 10 simulation scenes (S1-S10) with RDLUF-Mix$S^2$-9stg , RDLUF-Mix$S^2$-11stg, our 9stage, and our 11stage models.}
\label{table_22}
\renewcommand{\arraystretch}{1.5}.  %行宽
\setlength{\tabcolsep}{2.8pt}   %列宽
\centering
%\scriptsize
\footnotesize
%\tiny
\begin{tabular}{|c |c| c|c|c|c| c|c|c|c |c| c| c|c| c|c |c|c| c|c |c|c| c|c |c|c| c|} % {p{8cm} p{2.5cm}<{\centering}}
	\hline
	% \multirow{2}{*}{Name}，2为所占的行数，此语句可以使得内容垂直居中
	% \multicolumn{2}{c|}{Flag}，2为所占的列数，格式由第二个{}控制
	% \cline{2-3}指本行的2,3列画横线
	\multirow{1}{*}{IMP}   
	%&\multicolumn{1}{c|}{Params} 
	%&\multicolumn{1}{c|}{GFLOPs}  
	&\multicolumn{1}{c|}{S1}	 
	&\multicolumn{1}{c|}{S2} 
	
	&\multicolumn{1}{c|}{S3} 
	&\multicolumn{1}{c|}{S4}  
	&\multicolumn{1}{c|}{S5}	 
	%&\multicolumn{2}{c|}{GISCnet:32}
	
       &\multicolumn{1}{c|}{S6} 
	&\multicolumn{1}{c|}{S7}  
	&\multicolumn{1}{c|}{S8}	 
	&\multicolumn{1}{c|}{S9}	
	&\multicolumn{1}{c|}{S10}	
	&\multicolumn{1}{c|}{Avg}	
	
			\\ \cline{1-12}

 \multirow{2}{*}{RDLUF-MixS$^2$-9stg }    % &   \multirow{2}{*}{1.89M}  &  \multirow{2}{*}{231.09} 
 & 37.94  & 40.95 & 43.25  & 47.83  & 37.11  & 37.47  & 38.58  & 35.50  & 41.83  & 35.23 & \bf39.57 \\
    & 0.966  & 0.977  & 0.979  & 0.990  & 0.976  & 0.975  & 0.969 & 0.970  & 0.978  & 0.962  & \bf0.974	               			             
					             \\  \cline{1-12} 	
					             				              			        
   \multirow{2}{*}{RDLUF-MixS$^2$-11stg } % &   \multirow{2}{*}{1.89M}  &  \multirow{2}{*}{282.93} 	
 
 & 37.73  & 40.89 & 43.33  & 47.38  & 36.86  & 37.47  & 38.82  & 35.77  & 41.81  & 35.13 & \bf39.52 \\
  & 0.965  & 0.978  & 0.979 & 0.989  & 0.976  & 0.974  & 0.970 & 0.972  & 0.978  & 0.962  & \bf0.974

					              \\  \cline{1-12} 		
					              
 \multirow{2}{*}{Ours 9stage }     %  &   \multirow{2}{*}{2.46M}  &  \multirow{2}{*}{278.20} 
 & 37.90  & 41.22 & 43.55  & 47.87  & 37.25  & 37.62  & 38.92  & 36.28 & 42.53 & 35.06 & \bf39.82 \\
   & 0.966  & 0.979  & 0.980  & 0.990  & 0.977  & 0.975  & 0.969 & 0.973  & 0.980  & 0.961  & \bf0.975	   
                                                \\  \cline{1-12} 						              
					              				        
   \multirow{2}{*}{Ours 11stage}     %&   \multirow{2}{*}{2.46M}  &  \multirow{2}{*}{340.50} 
  &37.84  & 41.24 & 43.18  & 47.68  & 37.55  & 37.71  & 38.86 & 36.10  & 42.43 & 35.21 &  \bf39.78 \\
  & 0.966  & 0.980  & 0.979  & 0.989  & 0.979  & 0.976  & 0.970 & 0.974  & 0.980  & 0.962  &  \bf0.975
             
             \\  \hline    		                                   
\end{tabular}
\end{table*}

{ \bf Comparative Analysis of Network Reconstruction Performance Limits with RDLUF-Mix$S^2$.} The reconstructed performance of both PGDUDST and RDLUF-Mix$S^2$ \cite{rdluf} was further compared by increasing the stage number to 11 and training PGDUDST-11stg and RDLUF-Mix$S^2$-11stg.

The corresponding best reconstruction results for PGDUDST-11stg and RDLUF-Mix$S^2$-11stg are outlined in Table \ref{table_22}. A closer examination of Table \ref{table_22} reveals that the reconstruction outcomes of PGDUDST-11stg are marginally inferior to those of PGDUDST-9stg. Similarly, the reconstruction results of RDLUF-Mix$S^2$-11stg are somewhat worse than those of RDLUF-Mix$S^2$-9stg. This observation suggests that, irrespective of whether it is PGDUDST or RDLUF-Mix$S^2$ \cite{rdluf}, the reconstruction results do not unconditionally improve with an increase in the stage number; instead, they may experience some weakening. In essence, the reconstruction performance of both PGDUDST and RDLUF-Mix$S^2$ has its limitations. Referring to Table \ref{table_1} and Table \ref{table_22}, we can draw the following conclusions: PGDUDST-9stg and RDLUF-Mix$S^2$-9stg serve as the respective network reconstruction performance limits for PGDUDST and RDLUF-Mix$S^2$. Moreover, it is evident that the reconstruction limit of PGDUDST exceeds that of RDLUF-Mix$S^2$, implying that our proposed PGDUDST surpasses the network's reconstruction limit of RDLUF-Mix$S^2$.

{\bf Simulation HSI Reconstruction.}  We present a comparison of the proposed PGDUDST method for HSI reconstruction, utilizing 4 out of 28 spectral channels from Scene 2 and Scene 5, against the simulation results obtained from seven SOTA approaches.   Displayed in Fig. \ref{simu_spectral}, our method yields visually smoother and cleaner textures while retaining the spatial information of homogeneous regions. The results emphasize the effectiveness of our approach in producing high-quality HSIs with improved texture characteristics and preserved spatial information. Specifically, our method leverages the Spectral-attention branch to effectively model long-range dependencies and enhances the capability to capture detailed textures through the Dense-spatial branch. Additionally, we validate the spectral consistency of our approach by comparing the spectral density curves of the reconstructed areas with the actual ground truth.  As shown in the bottom-left of each scene in Fig. \ref{simu_spectral}, our method achieved the highest correlation coefficient, emphasizing the effectiveness of our proposed method.

   \begin{figure}[htpb]
\centering
\includegraphics[width=17.5cm,angle=0]{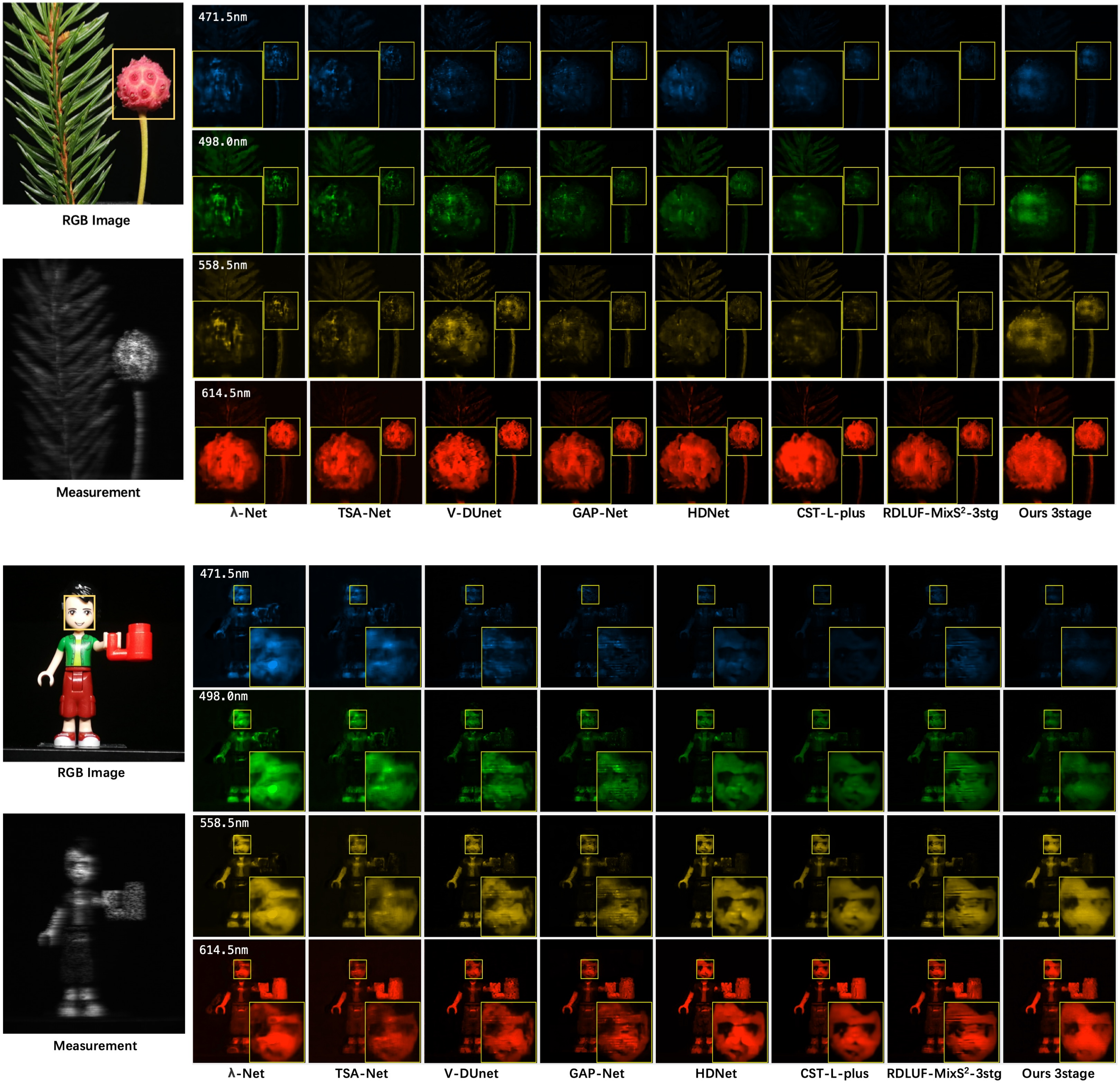}
\caption{Comparative Reconstruction of Real HSI: Scenes 2 and 3. Four spectra were randomly selected from a total of 28.}
\label{real_data}
\end{figure}

%In Scene 3, 我们方案出来的人脸是最为干净的(比如，不像其他方案那样会出来不存在的鼻子或者抬头纹，又或者出现很多其他伪影), 也是唯一能重建出代表嘴唇的两条曲线的，其他方案重建的嘴唇是无法分辨出来这两条曲线的。而且我们的重建方案的右眼也能看到更为清晰的眉毛。总的来说，从重建出来的重建来看，我们的方案大大降低了其他方案中存在的伪影问题，获得了最为干净和更加清晰的细节的重建的图像。 In Scene 3, the faces reconstructed by our approach exhibit exceptional clarity. Unlike other methods, our solution avoids generating non-existent features such as noses or forehead wrinkles and minimizes the presence of other artifacts. Notably, our approach is unique in accurately reconstructing the two distinct curves representing the lips, a detail that remains indiscernible in the reconstructions produced by alternative methods. Moreover, our reconstruction of the right eye provides a clearer depiction, offering enhanced visibility of the eyebrows. Overall, in terms of the reconstructed images, our approach markedly reduces the problem of artifacts present in other solutions, achieving reconstructions that are cleaner and more defined in detail.

{\bf Real HSI Reconstruction.} We retrained our model on the CAVE \cite{cave} and KAIST \cite{kaist} datasets, conducting tests on real measurements to implement actual experiments, following the configurations established in previous works \cite{mst, dgsmp, cst, tsa, rdluf}. To simulate real measurement conditions, we introduced 11-bit shot noise during the training process. As depicted in Fig. \ref{real_data}, we conducted a comparative analysis of the reconstructed images from two real scenes (Scene 2 and Scene 3) using our PGDUDST method and seven SOTA approaches. Our model achieved competitive results when compared to SOTA methods. Notably, in Scene 2, it is evident that the proposed method excels in restoring more texture and detail.  In Scene 3, the faces reconstructed by our approach exhibit exceptional clarity. Unlike other methods, our solution avoids generating non-existent features such as noses or forehead wrinkles and minimizes the presence of other artifacts. Notably, our approach is unique in accurately reconstructing the two distinct curves representing the lips, a detail that remains indiscernible in the reconstructions produced by alternative methods. Moreover, our reconstruction of the right eye provides a clearer depiction, offering enhanced visibility of the eyebrows. Overall, in terms of the reconstructed images, our approach markedly reduces the problem of artifacts present in other solutions, achieving reconstructions that are cleaner and more defined in detail.

%总之，本研究将密集空谱注意力变压器（DST）引入近端梯度下降展开框架（PGDUF），提出了一种名为近端梯度下降展开密集空谱注意力变压器（PGDUDST）的新方法，用于高光谱图像的重建。该方法有效解决了RDLUF-Mix$S^2$网络存在的两大缺陷问题：一是RDLUF-Mix$S^2$计算量巨大导致训练时间很长，二是RDLUF-Mix$S^2$在真实实验数据中表现不理想，真实数据的重建结果中包含大量伪影。PGDUDST则很好的解决了以上两个缺陷问题。首先，PGDUDST取得和RDLUF-Mix$S^2$-9stg相媲美的重建结果，但训练时间却只需要RDLUF-Mix$S^2$-9stg训练时长的58\%。其次，PGDUDST还超越了RDLUF-Mix$S^2$的重建极限，可以获得比RDLUF-Mix$S^2$更好的重建结果。最后，PGDUDST在真实实验数据上表现卓越，有效减轻了RDLUF-Mix$S^2$生成的重建图像中存在的众多伪影。这些结果突显了PGDUDST在高光谱图像重建中的高效性和有效性，展示了其在真实场景中加速收敛和提高适应性的潜力。所提出的方法有望推动高光谱成像技术在医学成像、遥感和目标跟踪等多个领域的实际应用。 

\section{Conclusion} 
In conclusion, this study introduces the Dense-spatial Spectral-attention Transformer (DST) into the Proximal Gradient Descent Unfolding Framework (PGDUF) and proposes a novel approach called Proximal Gradient Descent Unfolding Dense-spatial Spectral-attention Transformer (PGDUDST) for the reconstruction of hyperspectral images. This method effectively addresses two major flaws in the RDLUF-Mix$S^2$ network. First, the substantial computational burden of RDLUF-Mix$S^2$ results in excessively long training times. Second, RDLUF-Mix$S^2$ exhibits poor performance on real experimental data, leading to significant artifacts in the reconstructed results.

PGDUDST successfully overcomes these challenges. Firstly, it achieves reconstruction results comparable to RDLUF-Mix$S^2$-9stg but with only 58\% of the training time required by RDLUF-Mix$S^2$-9stg. Secondly, PGDUDST surpasses the reconstruction limits of RDLUF-Mix$S^2$, obtaining superior results. Lastly, PGDUDST demonstrates outstanding performance on real experimental data, effectively mitigating numerous artifacts present in the reconstructed images generated by RDLUF-Mix$S^2$. These results underscore the efficiency and effectiveness of PGDUDST in hyperspectral image reconstruction, showcasing its potential to accelerate convergence and improve adaptability in real-world scenarios. The proposed method holds promise for advancing the practical applications of hyperspectral imaging technology in various fields, including medical imaging, remote sensing, and target tracking.

}

{
    \small
    \bibliographystyle{ieeenat_fullname}
    \bibliography{CASSI}

\begin{thebibliography}{69}
\providecommand{\natexlab}[1]{#1}
\providecommand{\url}[1]{\texttt{#1}}
\expandafter\ifx\csname urlstyle\endcsname\relax
  \providecommand{\doi}[1]{doi: #1}\else
  \providecommand{\doi}{doi: \begingroup \urlstyle{rm}\Url}\fi

\bibitem[Arce et~al.(2014)Arce, Brady, Carin, Arguello, and
  Kittle]{arce2013compressive}
Gonzalo~R. Arce, David~J. Brady, Lawrence Carin, Henry Arguello, and David~S.
  Kittle.
\newblock Compressive coded aperture spectral imaging: An introduction.
\newblock \emph{IEEE Signal Processing Magazine}, 31\penalty0 (1):\penalty0
  105--115, 2014.

\bibitem[Beck and Teboulle(2009)]{beck2009fast}
Amir Beck and Marc Teboulle.
\newblock A fast iterative shrinkage-thresholding algorithm for linear inverse
  problems.
\newblock \emph{SIAM journal on imaging sciences}, 2\penalty0 (1):\penalty0
  183--202, 2009.

\bibitem[Bioucas-Dias and Figueiredo(2007)]{bioucas2007new}
JosÉ~M. Bioucas-Dias and MÁrio A.~T. Figueiredo.
\newblock A new twist: Two-step iterative shrinkage/thresholding algorithms for
  image restoration.
\newblock \emph{IEEE Transactions on Image Processing}, 16\penalty0
  (12):\penalty0 2992--3004, 2007.

\bibitem[Cai et~al.(2022{\natexlab{a}})Cai, Lin, Hu, Wang, Yuan, Zhang,
  Timofte, and Gool]{cst}
Yuanhao Cai, Jing Lin, Xiaowan Hu, Haoqian Wang, Xin Yuan, Yulun Zhang, Radu
  Timofte, and Luc~Van Gool.
\newblock Coarse-to-fine sparse transformer for hyperspectral image
  reconstruction.
\newblock In \emph{ECCV}, 2022{\natexlab{a}}.

\bibitem[Cai et~al.(2022{\natexlab{b}})Cai, Lin, Hu, Wang, Yuan, Zhang,
  Timofte, and Gool]{mst}
Yuanhao Cai, Jing Lin, Xiaowan Hu, Haoqian Wang, Xin Yuan, Yulun Zhang, Radu
  Timofte, and Luc~Van Gool.
\newblock Mask-guided spectral-wise transformer for efficient hyperspectral
  image reconstruction.
\newblock In \emph{CVPR}, 2022{\natexlab{b}}.

\bibitem[Cai et~al.(2022{\natexlab{c}})Cai, Lin, Lin, Wang, Zhang, Pfister,
  Timofte, and Van~Gool]{mst_plus_plus}
Yuanhao Cai, Jing Lin, Zudi Lin, Haoqian Wang, Yulun Zhang, Hanspeter Pfister,
  Radu Timofte, and Luc Van~Gool.
\newblock Mst++: Multi-stage spectral-wise transformer for efficient spectral
  reconstruction.
\newblock In \emph{Proceedings of the IEEE/CVF Conference on Computer Vision
  and Pattern Recognition (CVPR) Workshops}, pages 745--755,
  2022{\natexlab{c}}.

\bibitem[Cai et~al.(2022{\natexlab{d}})Cai, Lin, Wang, Yuan, Ding, Zhang,
  Timofte, and Gool]{dauhst}
Yuanhao Cai, Jing Lin, Haoqian Wang, Xin Yuan, Henghui Ding, Yulun Zhang, Radu
  Timofte, and Luc~Van Gool.
\newblock Degradation-aware unfolding half-shuffle transformer for spectral
  compressive imaging.
\newblock \emph{arXiv preprint arXiv:2205.10102}, 2022{\natexlab{d}}.

\bibitem[Cao et~al.(2016)Cao, Yue, Lin, Lin, Yuan, Dai, Carin, and
  Brady]{cao2016computational}
Xun Cao, Tao Yue, Xing Lin, Stephen Lin, Xin Yuan, Qionghai Dai, Lawrence
  Carin, and David~J Brady.
\newblock Computational snapshot multispectral cameras: Toward dynamic capture
  of the spectral world.
\newblock \emph{IEEE Signal Processing Magazine}, 33\penalty0 (5):\penalty0
  95--108, 2016.

\bibitem[Chan and Whiteman(1983)]{PSNR}
Luen~C. Chan and Peter Whiteman.
\newblock Hardware-constrained hybrid coding of video imagery.
\newblock \emph{IEEE Transactions on Aerospace and Electronic Systems},
  AES-19\penalty0 (1):\penalty0 71--84, 1983.

\bibitem[Chan et~al.(2016)Chan, Wang, and Elgendy]{chan2016plug}
Stanley~H Chan, Xiran Wang, and Omar~A Elgendy.
\newblock Plug-and-play admm for image restoration: Fixed-point convergence and
  applications.
\newblock \emph{IEEE Transactions on Computational Imaging}, 3\penalty0
  (1):\penalty0 84--98, 2016.

\bibitem[Chen et~al.(2022{\natexlab{a}})Chen, Wu, Wang, Hu, Hu, Ding, Cheng,
  and Wang]{chen2022mixformer}
Qiang Chen, Qiman Wu, Jian Wang, Qinghao Hu, Tao Hu, Errui Ding, Jian Cheng,
  and Jingdong Wang.
\newblock Mixformer: Mixing features across windows and dimensions.
\newblock In \emph{Proceedings of the IEEE/CVF conference on computer vision
  and pattern recognition}, pages 5249--5259, 2022{\natexlab{a}}.

\bibitem[Chen et~al.(2022{\natexlab{b}})Chen, Zhang, Gu, Liu, Suo, and
  Shao]{chen2022snapshot}
Wenwu Chen, Bo Zhang, Liuning Gu, Haibo Liu, Jinli Suo, and Xinxing Shao.
\newblock Snapshot compressive imaging based digital image correlation:
  temporally super-resolved full-resolution deformation measurement.
\newblock \emph{Optics Express}, 30\penalty0 (19):\penalty0 33554--33573,
  2022{\natexlab{b}}.

\bibitem[Chen et~al.(2022{\natexlab{c}})Chen, Liu, Hu, Wu, Wu, Lin, Tong, Yu,
  and Han]{czydgi}
Ziyan Chen, Zhentao Liu, Chenyu Hu, Heng Wu, Jianrong Wu, Jinda Lin, Zhishen
  Tong, Hong Yu, and Shensheng Han.
\newblock Hyperspectral image reconstruction for spectral camera based on ghost
  imaging via sparsity constraints using v-dunet.
\newblock \emph{arXiv preprint arXiv:2206.14199}, 2022{\natexlab{c}}.

\bibitem[Choi et~al.(2017)Choi, Kim, Gutierrez, Jeon, and Nam]{kaist}
Inchang Choi, MH Kim, D Gutierrez, DS Jeon, and G Nam.
\newblock High-quality hyperspectral reconstruction using a spectral prior.
\newblock Technical report, 2017.

\bibitem[Dong et~al.(2023)Dong, Gao, Qiu, Li, Yang, and Shi]{rdluf}
Yubo Dong, Dahua Gao, Tian Qiu, Yuyan Li, Minxi Yang, and Guangming Shi.
\newblock Residual degradation learning unfolding framework with mixing priors
  across spectral and spatial for compressive spectral imaging.
\newblock In \emph{Proceedings of the IEEE/CVF Conference on Computer Vision
  and Pattern Recognition}, pages 22262--22271, 2023.

\bibitem[Du et~al.(2009)Du, Tong, Cao, and Lin]{du2009prism}
Hao Du, Xin Tong, Xun Cao, and Stephen Lin.
\newblock A prism-based system for multispectral video acquisition.
\newblock In \emph{2009 IEEE 12th International Conference on Computer Vision},
  pages 175--182. IEEE, 2009.

\bibitem[Figueiredo et~al.(2007)Figueiredo, Nowak, and
  Wright]{figueiredo2007gradient}
M{\'a}rio~AT Figueiredo, Robert~D Nowak, and Stephen~J Wright.
\newblock Gradient projection for sparse reconstruction: Application to
  compressed sensing and other inverse problems.
\newblock \emph{IEEE Journal of selected topics in signal processing},
  1\penalty0 (4):\penalty0 586--597, 2007.

\bibitem[Fu et~al.(2021)Fu, Liang, and You]{fu2021bidirectional}
Ying Fu, Zhiyuan Liang, and Shaodi You.
\newblock Bidirectional 3d quasi-recurrent neural network for hyperspectral
  image super-resolution.
\newblock \emph{IEEE Journal of Selected Topics in Applied Earth Observations
  and Remote Sensing}, 14:\penalty0 2674--2688, 2021.

\bibitem[Gao and Smith(2015)]{gao2015optical}
Liang Gao and R~Theodore Smith.
\newblock Optical hyperspectral imaging in microscopy and spectroscopy--a
  review of data acquisition.
\newblock \emph{Journal of biophotonics}, 8\penalty0 (6):\penalty0 441--456,
  2015.

\bibitem[Gehm et~al.(2007)Gehm, John, Brady, Willett, and
  Schulz]{gehm2007single}
Michael~E Gehm, Renu John, David~J Brady, Rebecca~M Willett, and Timothy~J
  Schulz.
\newblock Single-shot compressive spectral imaging with a dual-disperser
  architecture.
\newblock \emph{Optics express}, 15\penalty0 (21):\penalty0 14013--14027, 2007.

\bibitem[Gu et~al.(2021)Gu, Liu, Gao, Ren, Ma, Chanussot, and
  Jia]{gu2021multimodal}
Yanfeng Gu, Tianzhu Liu, Guoming Gao, Guangbo Ren, Yi Ma, Jocelyn Chanussot,
  and Xiuping Jia.
\newblock Multimodal hyperspectral remote sensing: An overview and perspective.
\newblock \emph{Science China Information Sciences}, 64:\penalty0 1--24, 2021.

\bibitem[Hu et~al.(2018)Hu, Shen, and Sun]{hu2018squeeze}
Jie Hu, Li Shen, and Gang Sun.
\newblock Squeeze-and-excitation networks.
\newblock In \emph{Proceedings of the IEEE conference on computer vision and
  pattern recognition}, pages 7132--7141, 2018.

\bibitem[Hu et~al.(2022)Hu, Cai, Lin, Wang, Yuan, Zhang, Timofte, and
  Gool]{hdnet}
Xiaowan Hu, Yuanhao Cai, Jing Lin, Haoqian Wang, Xin Yuan, Yulun Zhang, Radu
  Timofte, and Luc~Van Gool.
\newblock Hdnet: High-resolution dual-domain learning for spectral compressive
  imaging.
\newblock In \emph{CVPR}, 2022.

\bibitem[Huang et~al.(2017)Huang, Liu, van~der Maaten, and
  Weinberger]{huang2017densely}
Gao Huang, Zhuang Liu, Laurens van~der Maaten, and Kilian~Q. Weinberger.
\newblock Densely connected convolutional networks.
\newblock In \emph{Proceedings of the IEEE Conference on Computer Vision and
  Pattern Recognition (CVPR)}, pages 4700--4708, 2017.

\bibitem[Huang et~al.(2021{\natexlab{a}})Huang, Dong, Yuan, Wu, and Shi]{dgsmp}
Tao Huang, Weisheng Dong, Xin Yuan, Jinjian Wu, and Guangming Shi.
\newblock Deep gaussian scale mixture prior for spectral compressive imaging.
\newblock In \emph{Proceedings of the IEEE/CVF Conference on Computer Vision
  and Pattern Recognition (CVPR)}, pages 16216--16225, 2021{\natexlab{a}}.

\bibitem[Huang et~al.(2021{\natexlab{b}})Huang, Dong, Yuan, Wu, and
  Shi]{huang2021deep}
Tao Huang, Weisheng Dong, Xin Yuan, Jinjian Wu, and Guangming Shi.
\newblock Deep gaussian scale mixture prior for spectral compressive imaging.
\newblock In \emph{Proceedings of the IEEE/CVF Conference on Computer Vision
  and Pattern Recognition}, pages 16216--16225, 2021{\natexlab{b}}.

\bibitem[Imani and Ghassemian(2020)]{imani2020overview}
Maryam Imani and Hassan Ghassemian.
\newblock An overview on spectral and spatial information fusion for
  hyperspectral image classification: Current trends and challenges.
\newblock \emph{Information fusion}, 59:\penalty0 59--83, 2020.

\bibitem[Karim et~al.(2023)Karim, Qadir, Farooq, Shakir, and
  Laghari]{HSImedicalimaging1}
Shahid Karim, Akeel Qadir, Umar Farooq, Muhammad Shakir, and Asif~A Laghari.
\newblock Hyperspectral imaging: a review and trends towards medical imaging.
\newblock \emph{Current Medical Imaging}, 19\penalty0 (5):\penalty0 417--427,
  2023.

\bibitem[Kittle et~al.(2010)Kittle, Choi, Wagadarikar, and
  Brady]{kittle2010multiframe}
David Kittle, Kerkil Choi, Ashwin Wagadarikar, and David~J Brady.
\newblock Multiframe image estimation for coded aperture snapshot spectral
  imagers.
\newblock \emph{Applied optics}, 49\penalty0 (36):\penalty0 6824--6833, 2010.

\bibitem[Laghari et~al.(2022)Laghari, Yin, et~al.]{HSImedicalimaging3}
Asif~Ali Laghari, S Yin, et~al.
\newblock How to collect and interpret medical pictures captured in highly
  challenging environments that range from nanoscale to hyperspectral imaging.
\newblock \emph{Current Medical Imaging}, 54:\penalty0 36582065, 2022.

\bibitem[Li et~al.(2020)Li, Xiong, Zhou, Wang, Lu, and Qian]{li2020bae}
Zhuanfeng Li, Fengchao Xiong, Jun Zhou, Jing Wang, Jianfeng Lu, and Yuntao
  Qian.
\newblock Bae-net: A band attention aware ensemble network for hyperspectral
  object tracking.
\newblock In \emph{2020 IEEE international conference on image processing
  (ICIP)}, pages 2106--2110. IEEE, 2020.

\bibitem[Liu et~al.(2018)Liu, Yuan, Suo, Brady, and Dai]{liu2018rank}
Yang Liu, Xin Yuan, Jinli Suo, David~J Brady, and Qionghai Dai.
\newblock Rank minimization for snapshot compressive imaging.
\newblock \emph{IEEE transactions on pattern analysis and machine
  intelligence}, 41\penalty0 (12):\penalty0 2990--3006, 2018.

\bibitem[Llull et~al.(2013)Llull, Liao, Yuan, Yang, Kittle, Carin, Sapiro, and
  Brady]{llull2013coded}
Patrick Llull, Xuejun Liao, Xin Yuan, Jianbo Yang, David Kittle, Lawrence
  Carin, Guillermo Sapiro, and David~J Brady.
\newblock Coded aperture compressive temporal imaging.
\newblock \emph{Optics express}, 21\penalty0 (9):\penalty0 10526--10545, 2013.

\bibitem[Lv and Wang(2020)]{HSIimageclassification1}
Wenjing Lv and Xiaofei Wang.
\newblock Overview of hyperspectral image classification.
\newblock \emph{Journal of Sensors}, 2020, 2020.

\bibitem[Ma et~al.(2019)Ma, Liu, Shou, and Yuan]{admm}
Jiawei Ma, Xiao-Yang Liu, Zheng Shou, and Xin Yuan.
\newblock Deep tensor admm-net for snapshot compressive imaging.
\newblock In \emph{Proceedings of the IEEE/CVF International Conference on
  Computer Vision (ICCV)}, 2019.

\bibitem[Meng et~al.(2020{\natexlab{a}})Meng, Jalali, and Yuan]{gap}
Ziyi Meng, Shirin Jalali, and Xin Yuan.
\newblock Gap-net for snapshot compressive imaging.
\newblock \emph{arXiv preprint arXiv:2012.08364}, 2020{\natexlab{a}}.

\bibitem[Meng et~al.(2020{\natexlab{b}})Meng, Ma, and Yuan]{tsa}
Ziyi Meng, Jiawei Ma, and Xin Yuan.
\newblock End-to-end low cost compressive spectral imaging with
  spatial-spectral self-attention.
\newblock In \emph{European conference on computer vision}, pages 187--204.
  Springer, 2020{\natexlab{b}}.

\bibitem[Meng et~al.(2020{\natexlab{c}})Meng, Qiao, Ma, Yu, Xu, and
  Yuan]{meng2020snapshot}
Ziyi Meng, Mu Qiao, Jiawei Ma, Zhenming Yu, Kun Xu, and Xin Yuan.
\newblock Snapshot multispectral endomicroscopy.
\newblock \emph{Optics Letters}, 45\penalty0 (14):\penalty0 3897--3900,
  2020{\natexlab{c}}.

\bibitem[Meng et~al.(2021)Meng, Yu, Xu, and Yuan]{meng2021self}
Ziyi Meng, Zhenming Yu, Kun Xu, and Xin Yuan.
\newblock Self-supervised neural networks for spectral snapshot compressive
  imaging.
\newblock In \emph{Proceedings of the IEEE/CVF international conference on
  computer vision}, pages 2622--2631, 2021.

\bibitem[Miao et~al.(2019)Miao, Yuan, Pu, and Athitsos]{lambda}
Xin Miao, Xin Yuan, Yunchen Pu, and Vassilis Athitsos.
\newblock l-net: Reconstruct hyperspectral images from a snapshot measurement.
\newblock In \emph{Proceedings of the IEEE/CVF International Conference on
  Computer Vision}, pages 4059--4069, 2019.

\bibitem[Mou et~al.(2022)Mou, Wang, and Zhang]{SPAN1}
Chong Mou, Qian Wang, and Jian Zhang.
\newblock Deep generalized unfolding networks for image restoration.
\newblock In \emph{Proceedings of the IEEE/CVF Conference on Computer Vision
  and Pattern Recognition (CVPR)}, pages 17399--17410, 2022.

\bibitem[Pande and Moharir(2023)]{pande2023application}
Chaitanya~B Pande and Kanak~N Moharir.
\newblock Application of hyperspectral remote sensing role in precision farming
  and sustainable agriculture under climate change: A review.
\newblock \emph{Climate Change Impacts on Natural Resources, Ecosystems and
  Agricultural Systems}, pages 503--520, 2023.

\bibitem[Park et~al.(2007)Park, Lee, Grossberg, and Nayar]{cave}
Jong-Il Park, Moon-Hyun Lee, Michael~D Grossberg, and Shree~K Nayar.
\newblock Multispectral imaging using multiplexed illumination.
\newblock In \emph{2007 IEEE 11th International Conference on Computer Vision},
  pages 1--8. IEEE, 2007.

\bibitem[Park et~al.(2019)Park, Liu, Wang, and Zhu]{SPAN2}
Taesung Park, Ming-Yu Liu, Ting-Chun Wang, and Jun-Yan Zhu.
\newblock Semantic image synthesis with spatially-adaptive normalization.
\newblock In \emph{Proceedings of the IEEE/CVF conference on computer vision
  and pattern recognition}, pages 2337--2346, 2019.

\bibitem[Peyghambari and Zhang(2021)]{peyghambari2021hyperspectral}
Sima Peyghambari and Yun Zhang.
\newblock Hyperspectral remote sensing in lithological mapping, mineral
  exploration, and environmental geology: an updated review.
\newblock \emph{Journal of Applied Remote Sensing}, 15\penalty0 (3):\penalty0
  031501--031501, 2021.

\bibitem[Qiao et~al.(2020)Qiao, Liu, and Yuan]{qiao2020snapshot}
Mu Qiao, Xuan Liu, and Xin Yuan.
\newblock Snapshot spatial--temporal compressive imaging.
\newblock \emph{Optics letters}, 45\penalty0 (7):\penalty0 1659--1662, 2020.

\bibitem[Szegedy et~al.(2016)Szegedy, Vanhoucke, Ioffe, Shlens, and
  Wojna]{szegedy2016rethinking}
Christian Szegedy, Vincent Vanhoucke, Sergey Ioffe, Jon Shlens, and Zbigniew
  Wojna.
\newblock Rethinking the inception architecture for computer vision.
\newblock In \emph{Proceedings of the IEEE conference on computer vision and
  pattern recognition}, pages 2818--2826, 2016.

\bibitem[Szegedy et~al.(2017)Szegedy, Ioffe, Vanhoucke, and
  Alemi]{szegedy2017inception}
Christian Szegedy, Sergey Ioffe, Vincent Vanhoucke, and Alexander Alemi.
\newblock Inception-v4, inception-resnet and the impact of residual connections
  on learning.
\newblock In \emph{Proceedings of the AAAI conference on artificial
  intelligence}, 2017.

\bibitem[Tan et~al.(2015)Tan, Ma, Rueda, Baron, and Arce]{tan2015compressive}
Jin Tan, Yanting Ma, Hoover Rueda, Dror Baron, and Gonzalo~R Arce.
\newblock Compressive hyperspectral imaging via approximate message passing.
\newblock \emph{IEEE Journal of Selected Topics in Signal Processing},
  10\penalty0 (2):\penalty0 389--401, 2015.

\bibitem[Tang et~al.(2022)Tang, Liu, Ji, and Huang]{tang2022robust}
Yiming Tang, Yufei Liu, Ling Ji, and Hong Huang.
\newblock Robust hyperspectral object tracking by exploiting background-aware
  spectral information with band selection network.
\newblock \emph{IEEE Geoscience and Remote Sensing Letters}, 19:\penalty0 1--5,
  2022.

\bibitem[Terentev et~al.(2022)Terentev, Dolzhenko, Fedotov, and
  Eremenko]{terentev2022current}
Anton Terentev, Viktor Dolzhenko, Alexander Fedotov, and Danila Eremenko.
\newblock Current state of hyperspectral remote sensing for early plant disease
  detection: A review.
\newblock \emph{Sensors}, 22\penalty0 (3):\penalty0 757, 2022.

\bibitem[ul~Rehman and Qureshi(2021)]{HSImedicalimaging2}
Aziz ul Rehman and Shahzad~Ahmad Qureshi.
\newblock A review of the medical hyperspectral imaging systems and unmixing
  algorithms’ in biological tissues.
\newblock \emph{Photodiagnosis and Photodynamic Therapy}, 33:\penalty0 102165,
  2021.

\bibitem[Wagadarikar et~al.(2008)Wagadarikar, John, Willett, and
  Brady]{wagadarikar2008single}
Ashwin Wagadarikar, Renu John, Rebecca Willett, and David Brady.
\newblock Single disperser design for coded aperture snapshot spectral imaging.
\newblock \emph{Applied optics}, 47\penalty0 (10):\penalty0 B44--B51, 2008.

\bibitem[Wagadarikar et~al.(2009)Wagadarikar, Pitsianis, Sun, and
  Brady]{wagadarikar2009video}
Ashwin~A Wagadarikar, Nikos~P Pitsianis, Xiaobai Sun, and David~J Brady.
\newblock Video rate spectral imaging using a coded aperture snapshot spectral
  imager.
\newblock \emph{Optics express}, 17\penalty0 (8):\penalty0 6368--6388, 2009.

\bibitem[Wambugu et~al.(2021)Wambugu, Chen, Xiao, Tan, Wei, Liu, and
  Li]{wambugu2021hyperspectral}
Naftaly Wambugu, Yiping Chen, Zhenlong Xiao, Kun Tan, Mingqiang Wei, Xiaoxue
  Liu, and Jonathan Li.
\newblock Hyperspectral image classification on insufficient-sample and feature
  learning using deep neural networks: A review.
\newblock \emph{International Journal of Applied Earth Observation and
  Geoinformation}, 105:\penalty0 102603, 2021.

\bibitem[Wang et~al.(2015)Wang, Xiong, Gao, Shi, Zeng, and Wu]{wang2015high}
Lizhi Wang, Zhiwei Xiong, Dahua Gao, Guangming Shi, Wenjun Zeng, and Feng Wu.
\newblock High-speed hyperspectral video acquisition with a dual-camera
  architecture.
\newblock In \emph{Proceedings of the IEEE Conference on Computer Vision and
  Pattern Recognition}, pages 4942--4950, 2015.

\bibitem[Wang et~al.(2016)Wang, Xiong, Shi, Wu, and Zeng]{wang2016adaptive}
Lizhi Wang, Zhiwei Xiong, Guangming Shi, Feng Wu, and Wenjun Zeng.
\newblock Adaptive nonlocal sparse representation for dual-camera compressive
  hyperspectral imaging.
\newblock \emph{IEEE transactions on pattern analysis and machine
  intelligence}, 39\penalty0 (10):\penalty0 2104--2111, 2016.

\bibitem[Wang et~al.(2019)Wang, Sun, Fu, Kim, and Huang]{wang2019hyperspectral}
Lizhi Wang, Chen Sun, Ying Fu, Min~H Kim, and Hua Huang.
\newblock Hyperspectral image reconstruction using a deep spatial-spectral
  prior.
\newblock In \emph{Proceedings of the IEEE/CVF Conference on Computer Vision
  and Pattern Recognition}, pages 8032--8041, 2019.

\bibitem[Wang et~al.(2020)Wang, Sun, Zhang, Fu, and Huang]{wang2020dnu}
Lizhi Wang, Chen Sun, Maoqing Zhang, Ying Fu, and Hua Huang.
\newblock Dnu: Deep non-local unrolling for computational spectral imaging.
\newblock In \emph{Proceedings of the IEEE/CVF Conference on Computer Vision
  and Pattern Recognition}, pages 1661--1671, 2020.

\bibitem[Wang et~al.(2004)Wang, Bovik, Sheikh, and Simoncelli]{SSIM}
Zhou Wang, A.C. Bovik, H.R. Sheikh, and E.P. Simoncelli.
\newblock Image quality assessment: from error visibility to structural
  similarity.
\newblock \emph{IEEE Transactions on Image Processing}, 13\penalty0
  (4):\penalty0 600--612, 2004.

\bibitem[Xiong et~al.(2020)Xiong, Zhou, and Qian]{xiong2020material}
Fengchao Xiong, Jun Zhou, and Yuntao Qian.
\newblock Material based object tracking in hyperspectral videos.
\newblock \emph{IEEE Transactions on Image Processing}, 29:\penalty0
  3719--3733, 2020.

\bibitem[Yuan(2016)]{yuan2016generalized}
Xin Yuan.
\newblock Generalized alternating projection based total variation minimization
  for compressive sensing.
\newblock In \emph{2016 IEEE International conference on image processing
  (ICIP)}, pages 2539--2543. IEEE, 2016.

\bibitem[Yuan et~al.(2020)Yuan, Liu, Suo, and Dai]{yuan2020plug}
Xin Yuan, Yang Liu, Jinli Suo, and Qionghai Dai.
\newblock Plug-and-play algorithms for large-scale snapshot compressive
  imaging.
\newblock In \emph{Proceedings of the IEEE/CVF Conference on Computer Vision
  and Pattern Recognition}, pages 1447--1457, 2020.

\bibitem[Yuan et~al.(2021{\natexlab{a}})Yuan, Brady, and
  Katsaggelos]{yuan2021snapshot}
Xin Yuan, David~J. Brady, and Aggelos~K. Katsaggelos.
\newblock Snapshot compressive imaging: Theory, algorithms, and applications.
\newblock \emph{IEEE Signal Processing Magazine}, 38\penalty0 (2):\penalty0
  65--88, 2021{\natexlab{a}}.

\bibitem[Yuan et~al.(2021{\natexlab{b}})Yuan, Liu, Suo, Durand, and
  Dai]{yuan2021plug}
Xin Yuan, Yang Liu, Jinli Suo, Fredo Durand, and Qionghai Dai.
\newblock Plug-and-play algorithms for video snapshot compressive imaging.
\newblock \emph{IEEE Transactions on Pattern Analysis and Machine
  Intelligence}, 44\penalty0 (10):\penalty0 7093--7111, 2021{\natexlab{b}}.

\bibitem[Zamir et~al.(2022)Zamir, Arora, Khan, Hayat, Khan, and
  Yang]{zamir2022restormer}
Syed~Waqas Zamir, Aditya Arora, Salman Khan, Munawar Hayat, Fahad~Shahbaz Khan,
  and Ming-Hsuan Yang.
\newblock Restormer: Efficient transformer for high-resolution image
  restoration.
\newblock In \emph{Proceedings of the IEEE/CVF conference on computer vision
  and pattern recognition}, pages 5728--5739, 2022.

\bibitem[Zhang et~al.(2019)Zhang, Wang, Fu, Zhong, and
  Huang]{zhang2019computational}
Shipeng Zhang, Lizhi Wang, Ying Fu, Xiaoming Zhong, and Hua Huang.
\newblock Computational hyperspectral imaging based on dimension-discriminative
  low-rank tensor recovery.
\newblock In \emph{Proceedings of the IEEE/CVF International Conference on
  Computer Vision}, pages 10183--10192, 2019.

\bibitem[Zhang et~al.(2022)Zhang, Zhang, Xiong, Sun, and
  Zhang]{zhang2022herosnet}
Xuanyu Zhang, Yongbing Zhang, Ruiqin Xiong, Qilin Sun, and Jian Zhang.
\newblock Herosnet: Hyperspectral explicable reconstruction and optimal
  sampling deep network for snapshot compressive imaging.
\newblock In \emph{Proceedings of the IEEE/CVF Conference on Computer Vision
  and Pattern Recognition}, pages 17532--17541, 2022.

\bibitem[Zhao et~al.(2023)Zhao, Liu, Su, Xu, Yan, and Feng]{zhao2023tmtnet}
Chunhui Zhao, Hongjiao Liu, Nan Su, Congan Xu, Yiming Yan, and Shou Feng.
\newblock Tmtnet: A transformer-based multimodality information transfer
  network for hyperspectral object tracking.
\newblock \emph{Remote Sensing}, 15\penalty0 (4):\penalty0 1107, 2023.

\end{thebibliography}
}


\end{document}